\documentclass[preprint,showpacs,preprintnumbers,amsmath,amssymb]{revtex4}

\usepackage{graphicx}% Include figure files
\usepackage{dcolumn}% Align table columns on decimal point
\usepackage{bm}% bold math

\begin{document}

\preprint{PRD 78, 056005 (2008)}

\title{Some mass relations for mesons and baryons in Regge phenomenology}

\author{Xin-Heng Guo\footnote{e-mail: xhguo@bnu.edu.cn}}
\affiliation{\scriptsize{College of Nuclear Science and
Technology, Beijing Normal University, Beijing 100875, China}}

\author{Ke-Wei Wei\footnote{Corresponding author, e-mail:weikw@brc.bnu.edu.cn}}
\affiliation{\scriptsize{College of Nuclear Science and
Technology, Beijing Normal University, Beijing 100875, China}}

\author{Xing-Hua Wu\footnote{e-mail: singhwa.wu@gmail.com}}
\affiliation{\scriptsize{College of Nuclear Science and Technology, Beijing Normal University, Beijing 100875, China;\\
College of Physics and Information Engineering, Henan Normal
University, Xinxiang 453007, China}}

\date{\today\\}

\begin{abstract}
   In the quasilinear Regge trajectory ansatz,
some useful linear mass inequalities, quadratic mass inequalities
and quadratic mass equalities are derived for mesons and baryons.
   Based on these relations, mass ranges of some mesons and baryons are given.
  The masses of $\bar{b}c$ and $s\bar{s}$ belonging to the pseudoscalar, vector and tensor meson multiplets are also extracted.
  The $J^P$ of the baryon $\Xi_{cc}^+(3520)$ is assigned to be $\frac{1}{2}^+$.
  The numerical values for Regge slopes and intercepts of the $\frac{1}{2}^+$ and $\frac{3}{2}^+$ $SU(4)$ baryon trajectories are extracted and
the masses of the orbital excited baryons lying on the
$\frac{1}{2}^+$ and $\frac{3}{2}^+$ trajectories are estimated.
   The $J^P$ assignments of baryons $\Xi_c(2980)$, $\Xi_c (3055)$,
$\Xi_c(3077)$ and $\Xi_c (3123)$ are discussed.
 The predictions are in reasonable
agreement with the existing experimental data and those suggested
in many other different approaches.
 The mass relations and the predictions may
be useful for the discovery of the unobserved meson and baryon
states and the $J^P$ assignment of these states.

\end{abstract}

   %  12.40.Yx, 12.20.Lq, 12.10.Kt,  %%12.39.-x
                                                % PACS, the Physics and Astronomy  % Classification Scheme.
%\keywords{Suggested keywords}  %Use showkeys class option if keyword  display desired

\maketitle

\newpage

\section{Introduction}

  The study of hadronic physics has been a subject of intense interest.
 There are many hadronic states reported in recent years:
 $B_2^*$ \cite{B2-D0-CDF},
 $B_{s2}^*$ \cite{Bs2-D0-CDF},
 $\Xi_{cc}^+(3520)$ \cite{Xi_cc},
 $\Lambda_c^+(2880)$ \cite{Lambda_c2880-CLEO,Lambda_c2880-2940a,Lambda_c2880-2940b},
 $\Lambda_c^+(2940)$ \cite{Lambda_c2880-2940a,Lambda_c2880-2940b},
 $\Xi_c^{0,+}(2980, 3077)$ \cite{Xi_2980-Xi_3077a, Xi_2980-Xi_3077b},
 $\Xi_c^+(3055, 3123)$ \cite{xi_3055-3123-babar},
 $\Sigma_b^{(\ast)\pm}$ \cite{Sigma-b}
 and $\Xi_{b}^-$ \cite{Xi-b}.
   More and more states will be discovered in the near future.
   However, the properties of some states such as $\Xi_{cc}^+(3520)$ are still not very clear.
$\Xi_{cc}^+(3520)$ was reported as the doubly charmed baryon state by SELEX in two different decay modes \cite{Xi_cc}, but the $J^P$ number has not been determined.
 Moreover, it has not been confirmed by other experiments (notably by BABAR \cite{Xi_cc(3520)-BABAR}, BELLE \cite{Xi_cc(3520)-BELLE} and FOCUS \cite{Xi_cc(3520)-FOCUS}).
 According to the Particle Data Group's ``Review of Particle
Physics'' in 2006 \cite{PDG2006},
 many hadrons, especially heavy hadrons,
are still absent from the summary tables.
 Obviously, there is still a lot of work to be done both theoretically and experimentally.

   The eightfold way and the standard SU(3) Gell-Mann--Okubo (GMO) formula \cite{Gell-Mann--Okubo}
have played an important role in the historical progress in particle physics.
   However, the direct generalization of the GMO formula to the charmed and bottom hadrons cannot
agree well with experimental data due to higher-order breaking effects.
   Consequently, there are many works focused on the mass relations,
including inequalities \cite{inequalities-history, inequalities-QCD1,inequalities-QCD2,inequalities-QCD3}
and equalities \cite{De-minLi2004,Burakovsky-meson-relation,Burakovsky1997,decuplet-splitting,decuplet-splitting2, equalities-history,instanton-model, Hendry-Lichtenberg-quarkmodel,Verma-Khanna-SU4,Verma-Khanna-SU8,Singh-Verma-Khanna,Singh-Khanna,Singh, Tait-1973, sextet-Jenkins, Okubo-2-order}.  % in many different models.

 Quantum chromodynamics (QCD) has been
verified as an appropriate theory to describe strong interaction at
short distances. However, the application of QCD to the processes of
hadronic interactions at large distances is still limited by the
unsolved confinement problem.
  Nowadays calculations of hadronic properties, which are related to the nonperturbative effects,
are frequently carried out with the help of phenomenological models.
  Regge phenomenology (which was derived from the analysis of the properties of the scattering amplitude
in the complex angular momentum plane \cite{Regge-theory})
is one of the simplest ones among these phenomenological models.
  Regge theory is concerned with almost all aspects of strong interactions, including the particle spectra, the forces
between particles, and the high energy behavior of scattering amplitudes \cite{regge-book}.
 The quasilinear Regge trajectory ansatz, which is one of the most
effective and popular approaches for studying hadron spectra, can (at least at present)
give a reasonable description for the hadron spectroscopy
\cite{anisovich,De-minLi2004,Ailin Zhang,Burakovsky-meson-relation,Burakovsky1997},
although some suggestions that the realistic Regge trajectories could be nonlinear exist \cite{nonlinear}.

 As pointed out in Refs. \cite{De-minLi2004,Burakovsky-para-relation},
 Regge intercepts and slopes are useful for many spectral and
nonspectral purposes, for example, in the recombination
\cite{reson1} and fragmentation \cite{reson2} models.   Therefore,
as pointed out in Ref. \cite{Basdevant1985}, the slopes and
intercepts of the Regge trajectories are fundamental constants of
hadron dynamics, perhaps in general more important than the masses of
 particular states.
    Thus, the determination of Regge slopes and intercepts of hadrons is of great importance
 since this provides opportunities for a better understanding of the dynamics of strong interactions \cite{Burakovsky-para-relation}.
% in the processes of production of hadrons at high energies and of their production rates \cite{Burakovsky-para-relation}.

  In the quasilinear Regge trajectory ansatz, the numerical
values of the parameters of the Regge trajectories were extracted for mesons of different flavors
\cite{anisovich,De-minLi2004, Burakovsky-meson-relation,Kaidalov1982,Ailin Zhang}. % for different meson multiplets %by Li \emph{et al}
   Under the approximation that mesons or baryons in the light quark sector have the common Regge slopes,
Burakovsky \emph{et al.} derived two 6th power and one 14th power meson mass relations in Ref. \cite{Burakovsky-meson-relation},
and derived some new quadratic Gell-Mann--Okubo--type baryon mass equalities in Ref. \cite{Burakovsky1997}.
%and two 6th power baryon mass relations in Ref. \cite{Burakovsky1997}.
%   New quadratic Gell-Mann--Okubo--type baryon mass equalities were also derived in Ref. \cite{Burakovsky1997}.
   Using those new quadratic baryon mass relations % and those 6th power approximate baryon mass relations,
they predicted the masses of $\frac{1}{2}^+$ and $\frac{3}{2}^+$ charmed baryon states absent from the baryon summary table then.
(Here and below, $\frac{1}{2}^+$ and $\frac{3}{2}^+$ multiplets refer to the ground multiplets in which the total orbital angular momenta $L$=0.)  %total quark orbital angular momenta
    However, the numerical values for the parameters of the charmed baryon
Regge trajectories were not given in Ref. \cite{Burakovsky1997}.
%However, duo to the lack of the input data needed to get reliable results,
% in which the orbital angular momenta between the quark pairs are zero),

   In the present work, under the assumption that the quasilinear Regge trajectory ansatz is suitable to
describe meson spectra and baryon spectra with the requirements of the additivity of intercepts and inverse slopes,
the relations between slope ratios and masses of hadrons with different flavors
and the mass relations among hadrons will be studied.
    We will show that the linear mass GMO formula is virtually an inequality and the quadratic mass GMO formula is also an inequality with the sign opposite to the linear case.
   We will get a high-power mass equation which is very useful to predict the masses of $\bar{b}c$ states and the masses of pure $s\bar{s}$ states.
   We will also get some useful quadratic mass equations for baryons.
  The $J^P$ assignment of $\Xi_{cc}^+(3520)$, $\Xi_c(2980)$, $\Xi_c (3055)$,
$\Xi_c(3077)$ and $\Xi_c (3123)$ baryons will be discussed.
  The numerical values for the parameters of the $\frac{1}{2}^+$ and $\frac{3}{2}^+$  trajectories will be extracted and
the masses of the baryon states lying on the $\frac{1}{2}^+$ and $\frac{3}{2}^+$
trajectories will be estimated.

   The remainder of this paper is organized as follows.
   In Sec. II we briefly introduce the quasilinear Regge trajectory ansatz.
   Then, we extract the mass inequalities and mass equalities for mesons and baryons.
    In Sec. III we present some applications of the relations derived in Sec. II
 and discuss the $J^P$ assignment of $\Xi_{cc}^+(3520)$, $\Xi_c(2980)$, $\Xi_c (3055)$,
 $\Xi_c(3077)$ and $\Xi_c (3123)$ baryons.
  The parameters of the $\frac{1}{2}^+$ and $\frac{3}{2}^+$  trajectories are extracted and
the masses of the baryon states lying on the $\frac{1}{2}^+$ and $\frac{3}{2}^+$
trajectories are estimated.
  Finally, we give a discussion and conclusion in Sec. IV.
%A short discussion is given in Sec. IV.

\section{Framework}

%  In a series of recent papers \cite{anisovich,De-minLi2004,Burakovsky1997},
% it is indicated that the quasilinear Regge trajectory ansatz can, at least at present,
% give a reasonable description for the hadron spectroscopy,
% and the predictions may be useful for the discovery of new hadron states which have not been observed yet.

  It is known from Regge theory that all mesons and baryons are associated with Regge poles
which move in the complex angular momentum plane as a function of energy.
  The trajectory of a particular pole (Regge trajectory) is characterized by a set of internal quantum numbers
(baryon number $\mathcal{B}$, intrinsic parity $P$, strangeness $\mathcal{S}$, charmness $C$, bottomness $B$, etc.)
and by the evenness or oddness of the total spin $J$ for mesons ($J-\frac{1}{2}$ for baryons) \cite{Chew-Frautschi}.
% A family is characterized by a set of internal quantum numbers % (radial quantum number $\mathcal{N}$, baryon number $\mathcal{B}$, intrinsic parity $P$, strangeness $S$, charm $C$, bottomness $B$, etc.) % and by the evenness or oddness of physical $J$ for mesons or $J-\frac{1}{2}$ for baryons.
  The plots of Regge trajectories of hadrons in the $(J,M^2)$ plane are usually called Chew-Frautschi plots
(where $J$ and $M$ are respectively the total spins and the masses of the hadrons).
  In Fig. 1, we draw the Chew-Frautschi plots for some meson and baryon Regge trajectories.

\begin{figure}
%\begin{minipage}[c]{0.5\textwidth}
\centering \includegraphics[width=6in] {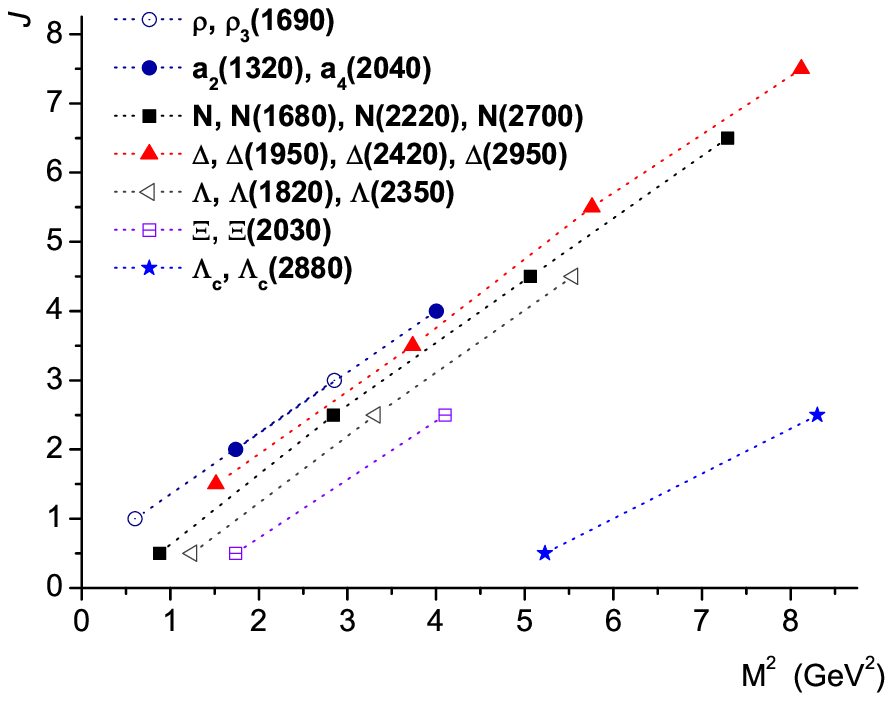}
%\end{minipage}\label{fig1}
\centerline{Figure 1. Chew-Frautschi plots in the ($J, M^2$) plane for some mesons and baryons.}
\end{figure}

   Assuming the existence of the quasilinear Regge trajectories for both light and heavy hadrons,
 one can have
% in the $(J,M^2)$ planes (Chew-Frautschi plots),
\begin{equation}
J=\alpha(M)=a(0)+\alpha^\prime M^2, \label{regge1}
\end{equation}
where $a(0)$ and $\alpha^\prime$ are respectively the
intercept and slope of the trajectory on which the particles lie.
  Hadrons lying on the same Regge trajectory which have the same internal quantum numbers
are classified into the same family.
% Each family lies on a given Regge trajectory with the value of total spin $J$ spaced by 2 units of angular momentum.
   The difference between the total spins of these hadrons is $2n$ ($n$=1,2,3,$\cdots$),
e.g., mesons with the quantum numbers $\mathcal{N} ^{2S+1}L_{J}$,
$\mathcal{N} ^{2S+1}(L+2)_{J+2}$, $\mathcal{N} ^{2S+1}(L+4)_{J+4}$, $\cdots$
(where $\mathcal{N}$, $L$ and $S$ denote the radial excited quantum number, % principal quantum number,
the orbital quantum number and the intrinsic spin, respectively)
lying on the same Regge trajectory.
   These features can be seen  from the well-known Chew-Frautschi plots (Fig. 1).

     For a meson multiplet with spin-parity $J^P$
(more exactly speaking, with quantum numbers $\mathcal{N} ^{2S+1}L_{J}$),   %spin-parity $J^P$,
the parameters for different quark constituents can be related by the following relations:

   the additivity of intercepts \cite{De-minLi2004,Burakovsky-meson-relation,Burakovsky-para-relation, dual-resonance, two-dimensional,dual-analytic, bremsstrahlung,inequalities-arguement,Kaidalov1982},
\begin{equation}
a_{i\bar{i}}(0)+a_{j\bar{j}}(0)=2a_{i\bar{j}}(0) \label{intercept1},
\end{equation}

   the additivity of inverse slopes \cite{De-minLi2004,Burakovsky-meson-relation,Burakovsky-para-relation, Kaidalov1982},
\begin{equation}
\frac{1}{\alpha_{i\bar{i}}^\prime}+\frac{1}{\alpha_{j\bar{j}}^\prime}=\frac{2}{\alpha_{i\overline{j}}^\prime}, \label{slope1a}
\end{equation}
 where $i$ and $j$ represent quark flavors.
% where $i$ and $j$ represent quarks with any flavor.
% where $i$ and $j$ represent the corresponding flavor contents.
   Equations (\ref{intercept1}) and (\ref{slope1a}) were derived in a model
 based on the topological expansion and the $q\bar{q}$-string picture of hadrons \cite{Kaidalov1982}.
   This model provides a microscopic approach to describe Regge phenomenology in terms of quark degrees of freedom \cite{Cassing-quark-degree}.
% which is the starting work of the Quark-Gluon String Model and can be considered as a microscopic model describing
   In fact, Eq. (\ref{intercept1}) was first derived for light quarks
in the dual-resonance model \cite{dual-resonance},
and was found to be satisfied in two-dimensional QCD \cite{two-dimensional},
the dual-analytic model \cite{dual-analytic}, and the quark
bremsstrahlung model \cite{bremsstrahlung}.
  Also, it saturates the inequality for Regge intercepts \cite{inequalities-arguement}
which follows from the Schwarz inequality and the unitarity relation.
% which follows from the s-channel unitarity condition.  %which was obtained by using of the Schwarz inequality for the overlap function for the unitarity relation.
% The relation (\ref{slope1a}) was derived based on topological expansion and the $q\bar{q}$-string picture \cite{Kaidalov1982}.
 The above two relations are usually generalized to
the baryon case \cite{bremsstrahlung,Burakovsky1997,Burakovsky-para-relation}, in which one has
\begin{equation}
a_{iiq}(0)+a_{jjq}(0)=2a_{ijq}(0) \label{intercept2},
\end{equation}
\begin{equation}
\frac{1}{\alpha_{iiq}^\prime}+\frac{1}{\alpha_{jjq}^\prime}=\frac{2}{\alpha_{ijq}^\prime}, \label{slope1b}
\end{equation}
where $q$ represents a quark.

 There are also relations about the factorization of slopes for mesons \cite{residues,residues-baryon} and baryons \cite{residues-baryon}:
\begin{equation}
{\alpha_{i\bar{i}}^\prime}\cdot{\alpha_{j\bar{j}}^\prime}={\alpha_{i\bar{j}}^\prime}^2,\label{factorization-slope-m}
\end{equation}
\begin{equation}
{\alpha_{iiq}^\prime}\cdot{\alpha_{jjq}^\prime}={\alpha_{ijq}^\prime}^2,\label{factorization-slope-b}
\end{equation}
%for meson and baryon respectively
which follow from the factorization of residues of the $t$-channel poles.
   The paper by Burakovsky and Goldman \cite{Burakovsky-para-relation} showed that
only the additivity of inverse Regge slopes is consistent with the formal chiral and heavy quark limits for both mesons and baryons,
and that the factorization of Regge slopes, although consistent with the formal chiral limit, fails in the heavy quark limit.
  Besides, in Sec. III B, we will show that
 the high-power equation (\ref{high-equal-factor}) derived from the relations (\ref{regge1}), (\ref{intercept1}) and (\ref{factorization-slope-m})
 is not as good as the high-power equation (\ref{high-equal-m}) derived from the relations (\ref{regge1}), (\ref{intercept1}) and (\ref{slope1a})
 compared with the well-established meson multiplets.
   Therefore, we will use the relations (\ref{slope1a}) and (\ref{slope1b}) (the additivity of inverse slopes)
 rather than the relations (\ref{factorization-slope-m}) and (\ref{factorization-slope-b}) (the factorization of slopes) in this study.
   There are also studies about the relations between the ground state and its radial excited states \cite{anisovich,Li-Liu-2007,Gershtein-2006}
and there are suggestions that the radial excited states lie on daughter trajectories of the ground state \cite{regge-book}.
 However,  we do not discuss these relations in the present work.
 %  which may be related the daughter trajectory.
 %  However, since $cdots$, we do not discuss this them in the present work.

\subsection{Relations between slope ratios and hadron masses}
 For mesons, using Eqs. (\ref{regge1}) and (\ref{intercept1}), one obtains
\begin{equation}
\alpha_{i\bar{i}}'M_{i\bar{i}}^2+\alpha_{j\bar{j}}'M_{j\bar{j}}^2=2\alpha_{i\bar{j}}'M_{i\bar{j}}^2, \label{combination1}
\end{equation}
where the meson states $i\bar{i}$, $j\bar{j}$ and $i\bar{j}$ belong to the same $\mathcal{N} ^{2S+1}L_J$ multiplet.
  This relation can be reduced to the quadratic Gell-Mann--Okubo-type formula by assuming that all the slopes are independent of flavors ($\alpha_{i\bar{i}}'=\alpha_{i\bar{j}}'=\alpha_{j\bar{j}}'$).
 Combining the relations (\ref{slope1a}) and (\ref{combination1}), one can get two pairs of solutions. The first pair of solutions are
\begin{equation}
\begin{cases}
\begin{aligned}
 \frac{\alpha_{j\bar{j}}'}{\alpha_{i\bar{i}}'}&=\frac{1}{2M_{j\bar{j}}^2}\times[(4M_{i\bar{j}}^2-M_{i\bar{i}}^2-M_{j\bar{j}}^2)+\sqrt{(4M_{i\bar{j}}^2-M_{i\bar{i}}^2-M_{j\bar{j}}^2)^2-4M_{i\bar{i}}^2M_{j\bar{j}}^2} ],\label{solution1}\\
 \frac{\alpha_{i\bar{j}}'}{\alpha_{i\bar{i}}'}&=\frac{1}{4M_{i\bar{j}}^2}\times[(4M_{i\bar{j}}^2+M_{i\bar{i}}^2-M_{j\bar{j}}^2)+\sqrt{(4M_{i\bar{j}}^2-M_{i\bar{i}}^2-M_{j\bar{j}}^2)^2-4M_{i\bar{i}}^2M_{j\bar{j}}^2} ],
\end{aligned}
\end{cases}
\end{equation}
while the second pair of solutions are
\begin{equation}
\begin{cases}
\begin{aligned}
\frac{\alpha_{j\bar{j}}'}{\alpha_{i\bar{i}}'}&=\frac{1}{2M_{j\bar{j}}^2}\times[(4M_{i\bar{j}}^2-M_{i\bar{i}}^2-M_{j\bar{j}}^2)-\sqrt{(4M_{i\bar{j}}^2-M_{i\bar{i}}^2-M_{j\bar{j}}^2)^2-4M_{i\bar{i}}^2M_{j\bar{j}}^2} ],\\
\frac{\alpha_{i\bar{j}}'}{\alpha_{i\bar{i}}'}&=\frac{1}{4M_{i\bar{j}}^2}\times[(4M_{i\bar{j}}^2+M_{i\bar{i}}^2-M_{j\bar{j}}^2)-\sqrt{(4M_{i\bar{j}}^2-M_{i\bar{i}}^2-M_{j\bar{j}}^2)^2-4M_{i\bar{i}}^2M_{j\bar{j}}^2} ]. \label{solution11}
\end{aligned}
\end{cases}
\end{equation}

   From Eq. (\ref{regge1}), one has
\begin{equation}
\alpha'=\frac{(J+2)-J}{M_{J+2}^2-M_J^2}     \label{alpha} .
\end{equation}
  It is obvious that the Regge slope $\alpha'$ should be a single positive real number.
 Thus, $\frac{\alpha_{j\bar{j}}'}{\alpha_{i\bar{i}}'}$ should take only one value for a multiplet with certain $i$ and $j$.
% Therefore, one of Eqs. (\ref{solution1}) and (\ref{solution11}) should be discard.
  Since the relations (\ref{slope1a}) and (\ref{combination1}) are symmetric
under the exchange of the quark flavors $i$ and $j$,  % flavor indices $i$ and $j$,
we only consider the case in which quark masses satisfy $m_i < m_j$ for mesons here and after.
%to discuss which pair of the two pairs of solutions should be discard.

  From Eqs. (\ref{solution1}) and (\ref{solution11}),
we have the values of $\frac{\alpha_{c\bar{c}}'}{\alpha_{n\bar{n}}'}$ and $\frac{\alpha_{b\bar{b}}'}{\alpha_{n\bar{n}}'}$ ($n$ denotes $u$ or $d$ quark)
for the well-established multiplets.
  In the calculation, we do not consider the small mass splittings
%among isospin multiplet which are believed to be
caused by isospin breaking effects due to electromagnetic interaction.
  Here and below, all the masses of hadrons used in calculation are taken from PDG2006 \cite{PDG2006} except for the newly observed hadrons.
The results are shown in Table 1.

\begin{table}
 Table 1.   The values of $\frac{\alpha_{c\bar{c}}'}{\alpha_{n\bar{n}}'}$ and $\frac{\alpha_{b\bar{b}}'}{\alpha_{n\bar{n}}'}$ ($n$ denotes $u$ or $d$ quark)
obtained from Eqs. (\ref{solution1}) and (\ref{solution11}).\\
\begin{ruledtabular}

\begin{tabular}{c*{4}{l}}
                              &  $\mathcal{N}^{2S+1}L_{J}$  &(\ref{solution1}) &(\ref{solution11})  &   \\\hline

$\alpha_{c\bar{c}}'/\alpha_{n\bar{n}}'$ &1 $^{1}S_{0}$  &0.5636         &0.0038       &   \\
                                        &1 $^{1}P_{1}$  &0.5433         &0.2238       &\\
                                        &1 $^{3}S_{1}$  &0.4921         &0.1274       &\\
                                        &1 $^{3}P_{2}$  &0.5041         &0.2726       & \\
\\

$\alpha_{b\bar{b}}'/\alpha_{n\bar{n}}'$ &1 $^{1}S_{0}$  &0.2880         &0.0008     &  \\
                                        &1 $^{3}S_{1}$  &0.2361         &0.0290     &\\
                                        &1 $^{3}P_{2}$  &0.2562         &0.0690       & \\
\end{tabular}

\end{ruledtabular}
\end{table}

  The values of $\alpha_{n\bar{n}}'$ for light nonstrange meson trajectories of different multiplets
are in the range 0.7$-$0.9 GeV$^{-2}$ \cite{Burakovsky-meson-relation, anisovich, De-minLi2004, Kaidalov1982, YuSimonov}.
  The values of $\alpha_{c\bar{c}}'$ and $\alpha_{b\bar{b}}'$ for charmonium and bottomonium trajectories of
different multiplets are in the ranges 0.3$-$0.5 GeV$^{-2}$ and 0.18$-$0.25 GeV$^{-2}$, respectively \cite{Burakovsky-meson-relation, De-minLi2004, Gershtein-2006, Kaidalov1982}.
% In Ref. \cite{Burakovsky-meson-relation}, under the approximation that mesons (baryons)
% in the light quark sector have the common Regge slopes,
% $\alpha_{c\bar{c}}'\sim0.50GeV^{-2}$ and $\alpha_{\bar{b}}'\sim0.18GeV^{-2}$ are derived.
  Then, we have $\frac{\alpha_{c\bar{c}}'}{\alpha_{n\bar{n}}'} \sim 0.5$ and $\frac{\alpha_{b\bar{b}}'}{\alpha_{n\bar{n}}'} \sim 0.27$.
   From Table 1, one can see that the values of $\frac{\alpha_{c\bar{c}}'}{\alpha_{n\bar{n}}'}$ ($\frac{\alpha_{b\bar{b}}'}{\alpha_{n\bar{n}}'}$)
given by Eq. (\ref{solution1}) are approximately the same for different multiplets as they should to be.
   However, the values of $\frac{\alpha_{c\bar{c}}'}{\alpha_{n\bar{n}}'}$ ($\frac{\alpha_{b\bar{b}}'}{\alpha_{n\bar{n}}'}$)
given by Eq. (\ref{solution11}) are quite different for different multiplets.
   Furthermore, the values of $\frac{\alpha_{c\bar{c}}'}{\alpha_{n\bar{n}}'}$ and $\frac{\alpha_{b\bar{b}}'}{\alpha_{n\bar{n}}'}$ given by Eq. (\ref{solution11}) are too small to be accepted.
  Therefore, we take the first pair of solutions (Eq. (\ref{solution1})) and discard the second pair of solutions (Eq. (\ref{solution11})).

% The values of $\alpha_{b\bar{b}}'$ are too small to satisfy the inequalities for Regge slopes [ ].
% when the quark masses $m_i < m_j$, we discard the second pair of solutions (\ref{solution11}),  % because these solutions always gave unreasonable small values, e.g. for the ground pseudoscalar $1^{1}S_{0}$ states, when $M_{ii}=M_\pi$, $M_{jj}=M_{\eta_{b}(1S)}$, $M_{ij}=M_{B}$,  % the second pair of solutions give unacceptable values Regge slopes, %$\alpha_{b\bar{b}}' \sim 0.0005$ and $\alpha_{n\bar{b}}' \sim 0.001$.
%  Obviously, when $m_j < m_i$, the second pair of solutions is right, the first pair of solutions should be discard.

   For baryons, using Eqs. (\ref{regge1}) and (\ref{intercept2}), one obtains
\begin{equation}
\alpha_{iiq}'M_{iiq}^2+\alpha_{jjq}'M_{jjq}^2=2\alpha_{ijq}'M_{ijq}^2, \label{combination2}
\end{equation}
where $q$ denotes an arbitrary light or heavy quark.
 Combining the relations (\ref{slope1b}) and (\ref{combination2}), one can get two pairs of solutions,
%  Similar to mesons, for baryons, when the quark masses $m_j > m_i$, one can have  %   We generalize Eq. (\ref{solution1}) to the case of baryons.  Then we have (quark masses $m_j > m_i$),
\begin{equation}
\begin{cases}
\begin{aligned}
 \frac{\alpha_{jjq}'}{\alpha_{iiq}'}&=\frac{1}{2M_{jjq}^2}\times[(4M_{ijq}^2-M_{iiq}^2-M_{jjq}^2)+\sqrt{(4M_{ijq}^2-M_{iiq}^2-M_{jjq}^2)^2-4M_{iiq}^2M_{jjq}^2} ],\label{solution-b}\\
 \frac{\alpha_{ijq}'}{\alpha_{iiq}'}&=\frac{1}{4M_{ijq}^2}\times[(4M_{ijq}^2+M_{iiq}^2-M_{jjq}^2)+\sqrt{(4M_{ijq}^2-M_{iiq}^2-M_{jjq}^2)^2-4M_{iiq}^2M_{jjq}^2} ],
\end{aligned}
\end{cases}
\end{equation}
and
\begin{equation}
\begin{cases}
\begin{aligned}
 \frac{\alpha_{jjq}'}{\alpha_{iiq}'}&=\frac{1}{2M_{jjq}^2}\times[(4M_{ijq}^2-M_{iiq}^2-M_{jjq}^2)-\sqrt{(4M_{ijq}^2-M_{iiq}^2-M_{jjq}^2)^2-4M_{iiq}^2M_{jjq}^2} ],\label{solution-b2}\\
 \frac{\alpha_{ijq}'}{\alpha_{iiq}'}&=\frac{1}{4M_{ijq}^2}\times[(4M_{ijq}^2+M_{iiq}^2-M_{jjq}^2)-\sqrt{(4M_{ijq}^2-M_{iiq}^2-M_{jjq}^2)^2-4M_{iiq}^2M_{jjq}^2} ].
\end{aligned}
\end{cases}
\end{equation}

     From the Chew-Frautschi plots (Fig. 1), it is obvious that the Regge slope $\alpha'$ should be a single positive real number.
     Thus, $ \frac{\alpha_{jjq}'}{\alpha_{iiq}'}$ should take only one value for a multiplet with certain $i$, $j$ and $q$.  % one of Eqs. (\ref{solution-b}) and (\ref{solution-b2}) should be discard.
  Since the relations (\ref{slope1b}) and (\ref{combination2}) are symmetric
under the exchange of the quark flavors $i$ and $j$,  % flavor indices $i$ and $j$,
we only consider the case in which quark masses satisfy $m_i < m_j$ for baryons here and after.

   For the $\frac{1}{2}^+$ multiplet,
   when $i=n$, $j=s$, and $q=n$,
we have $M_{nnn}=M_{N(939)}$, $M_{nss}=M_{\Xi}$, and $M_{nns}^2=\frac{1}{4}(3M_{\Lambda}^2+M_{\Sigma}^2)$ \cite{Burakovsky1997}.
  Then, we have $\frac{\alpha_\Xi'}{\alpha_N'}=0.89$ from Eq. (\ref{solution-b}) and $\frac{\alpha_\Xi'}{\alpha_N'}=0.57$ from Eq. (\ref{solution-b2}).
   For the $\frac{3}{2}^+$ multiplet, when $i=n$, $j=s$, and $q=n$,
we have $M_{nnn}=M_\Delta$, $M_{nns}=M_{\Sigma^*}$, and $M_{nss}=M_{\Xi^*}$.
Then, we have $\frac{\alpha_{\Xi^*}'}{\alpha_\Delta'}=0.89$ from Eq. (\ref{solution-b}) and $\frac{\alpha_{\Xi^*}'}{\alpha_\Delta'}=0.72$ from Eq. (\ref{solution-b2}).
   Since the Regge trajectories of light baryons are approximately parallel,
the values of $\frac{\alpha_\Xi'}{\alpha_N'}$ and $\frac{\alpha_{\Xi^*}'}{\alpha_\Delta'}$ should be close to 1.
   Therefore, Eqs. (\ref{solution-b2}) should be discarded in the case of quark masses $m_i < m_j$.
   Furthermore,  Eqs. (\ref{solution-b}) and (\ref{solution-b2}) can be considered as
the generalization of Eqs. (\ref{solution1}) and (\ref{solution11}) respectively from the meson case to the baryon case.
   Therefore, we take Eq. (\ref{solution-b}) and discard Eq. (\ref{solution-b2}).

\subsection{High-power mass equalities }

    From Eqs. (\ref{solution1}) and (\ref{solution-b}), high-power mass equalities can be derived for mesons and baryons, respectively.
    For mesons, using
\begin{equation}
\frac{\alpha_{j\bar{j}}'}{\alpha_{i\bar{i}}'}=\frac{\alpha_{k\bar{k}}'}{\alpha_{i\bar{i}}'}\times\frac{\alpha_{j\bar{j}}'}{\alpha_{k\bar{k}}'},
\end{equation}
and Eq. (\ref{solution1}), when $m_i<m_j<m_k$, we have
%When $m_i<m_j<m_k$, from relations (\ref{solution1}), one can have,

\begin{equation}
\begin{split}
&\frac{(4M_{i\bar{j}}^2-M_{i\bar{i}}^2-M_{j\bar{j}}^2)+\sqrt{(4M_{i\bar{j}}^2-M_{i\bar{i}}^2-M_{j\bar{j}}^2)^2-4M_{i\bar{i}}^2M_{j\bar{j}}^2}}{2M_{j\bar{j}}^2}
\\  =& \frac{[(4M_{i\bar{k}}^2-M_{i\bar{i}}^2-M_{k\bar{k}}^2)+\sqrt{(4M_{i\bar{k}}^2-M_{i\bar{i}}^2-M_{k\bar{k}}^2)^2-4M_{i\bar{i}}^2M_{k\bar{k}}^2}]/{2M_{k\bar{k}}^2}}{[(4M_{j\bar{k}}^2-M_{j\bar{j}}^2-M_{k\bar{k}}^2)+\sqrt{(4M_{j\bar{k}}^2-M_{j\bar{j}}^2-M_{k\bar{k}}^2)^2-4M_{j\bar{j}}^2M_{k\bar{k}}^2}]/{2M_{k\bar{k}}^2}}  \label{high-equal-m}.
\end{split}
\end{equation}

For baryons, using
\begin{equation}
\frac{\alpha_{jjq}'}{\alpha_{iiq}'}=\frac{\alpha_{kkq}'}{\alpha_{iiq}'}\times\frac{\alpha_{jjq}'}{\alpha_{kkq}'} ,
\end{equation}
and Eq. (\ref{solution-b}), when $m_i<m_j<m_k$, we have
%where q denotes a arbitrary light or heavy quark. When $m_i<m_j<m_k$, from relations (\ref{solution-b}), one can have,
\begin{equation}
\begin{split}
& \frac{(4M_{ijq}^2-M_{iiq}^2-M_{jjq}^2)+\sqrt{(4M_{ijq}^2-M_{iiq}^2-M_{jjq}^2)^2-4M_{iiq}^2M_{jjq}^2}}{2M_{jjq}^2} \\
=& \frac{[(4M_{ikq}^2-M_{iiq}^2-M_{kkq}^2)+\sqrt{(4M_{ikq}^2-M_{iiq}^2-M_{kkq}^2)^2-4M_{iiq}^2M_{kkq}^2}]/{2M_{kkq}^2}}{[(4M_{jkq}^2-M_{jjq}^2-M_{kkq}^2)+\sqrt{(4M_{jkq}^2-M_{jjq}^2-M_{kkq}^2)^2-4M_{jjq}^2M_{kkq}^2}]/{2M_{kkq}^2}}  \label{equa-double} ,
\end{split}
\end{equation}
where $q$ denotes an arbitrary light or heavy quark.

  Relations (\ref{high-equal-m}) and (\ref{equa-double}) are the high-power mass equalities among one $J^P$ multiplet.
  They can be used to predict the masses of unobserved states.
  In Sec. III, we will apply Eq. (\ref{high-equal-m}) to predict
 the masses of $\bar{b}c$ meson states and the masses of the pure $s\bar{s}$ meson states.

\subsection{Linear mass inequalities and quadratic mass inequalities}

    From Eqs. (\ref{solution1}) and (\ref{solution-b}),
 two kinds of interesting inequalities can be derived for mesons and baryons, respectively.
    For mesons, as mentioned in the above discussion, $\alpha_{j\bar{j}}'$
and $\alpha_{i\bar{i}}'$ ought to be positive real numbers.
  Thus $\frac{\alpha_{j\bar{j}}'}{\alpha_{i\bar{i}}'}$ should also be a real number.
Then from Eq. (\ref{solution1}), we have
\begin{equation}
|4M_{i\bar{j}}^2-M_{i\bar{i}}^2-M_{j\bar{j}}^2| \geq 2M_{i\bar{i}} M_{j\bar{j}}.  \label{ine-linear-middle}
\end{equation}

   When $i=j$, $4M_{i\bar{j}}^2-M_{i\bar{i}}^2-M_{j\bar{j}}^2 \leq 0$ cannot be held;
when $i \neq j$, $4M_{i\bar{j}}^2-M_{i\bar{i}}^2-M_{j\bar{j}}^2 \leq 0$ can be easily ruled out by the data of the well-established meson multiplets.
  Therefore, $4M_{i\bar{j}}^2-M_{i\bar{i}}^2-M_{j\bar{j}}^2 \geq 0$.
  Thus, Eq. (\ref{ine-linear-middle}) can be written as the following:
\begin{equation}
4M_{i\bar{j}}^2-M_{i\bar{i}}^2-M_{j\bar{j}}^2 \geq 2M_{i\bar{i}} M_{j\bar{j}}.
\end{equation}
This relation can be simplified to
\begin{equation}
2M_{i\bar{j}} \geq M_{i\bar{i}}+M_{j\bar{j}}. \label{inequa1m0}
\end{equation}

   If $i=j$, $M_{i\bar{i}}=M_{i\bar{j}}=M_{j\bar{j}}$, then we have $2M_{i\bar{j}}=M_{i\bar{i}}+M_{j\bar{j}}$.
%\\
   On the other hand, if $2M_{i\bar{j}}=M_{i\bar{i}}+M_{j\bar{j}}$, using Eq. (\ref{solution1}), we have
\begin{equation}
\frac{\alpha_{j\bar{j}}'}{\alpha_{i\bar{i}}'}=\frac{M_{i\bar{i}}}{M_{j\bar{j}}}.  \label{jj/ii}
\end{equation}
%   Eq. (\ref{jj/ii}) is a general expression for hadrons belonging to one multiplet. Therefore,
   From the derivation of Eq. (\ref{jj/ii}), we can see that this equation is valid for hadrons belonging to the same multiplet.
   Since hadrons lying on the same Regge trajectory (which have the total angular momenta $J$, $J+2$, $J+4$, $\cdots$)
 have the same slope, we have
\begin{equation}
\frac{\alpha_{j\bar{j}}'}{\alpha_{i\bar{i}}'}=\frac{M_{i\bar{i},J}}{M_{j\bar{j},J}}=\frac{M_{i\bar{i},J+2}}{M_{j\bar{j},J+2}}.  \label{jj/ii-2}
\end{equation}
  From Eq. (\ref{alpha}), we have
\begin{equation}
\alpha'_{i\bar{i}}=\frac{2}{M_{i\bar{i},J+2}^2-M_{i\bar{i},J}^2}, \hspace{0.5cm}  \alpha'_{j\bar{j}}=\frac{2}{M_{j\bar{j},J+2}^2-M_{j\bar{j},J}^2}.   \label{alpha-ii}
\end{equation}
  Combining Eqs. (\ref{jj/ii-2}) and (\ref{alpha-ii}), we have
\begin{equation}
\frac{\alpha_{j\bar{j}}'}{\alpha_{i\bar{i}}'}
=\frac{M_{i\bar{i},J+2}+M_{i\bar{i},J}}{M_{j\bar{j},J+2}+M_{j\bar{j},J}} \times \frac{M_{i\bar{i},J+2}-M_{i\bar{i},J}}{M_{j\bar{j},J+2}-M_{j\bar{j},J}}
=\left(\frac{\alpha_{j\bar{j}}'}{\alpha_{i\bar{i}}'}\right)^2.
\end{equation}
  As mentioned before, the Regge slope $\alpha'$ is a positive real number.
 Therefore, $\frac{\alpha_{j\bar{j}}'}{\alpha_{i\bar{i}'}}=1$ when $2M_{i\bar{j}}=M_{i\bar{i}}+M_{j\bar{j}}$.
 Consequently we have
$M_{i\bar{i},J}=M_{j\bar{j},J}$ and $M_{i\bar{i},J+2}=M_{j\bar{j},J+2}$ from Eq. (\ref{jj/ii-2}).
   This leads to $i=j$ since the $i\bar{i}$ and $j\bar{j}$ states have the same $J^P$.

%  Mass is a very important property for hadrons.
% $i$ and $j$ represent quark flavors. If $i\neq j$, it can not hold for $M_{i\bar{i},J}=M_{j\bar{j},J}$ and $M_{i\bar{i},J+2}=M_{j\bar{j},J+2}$.
% Therefore, $i=j$.

 From the above analysis, we can conclude that
if and only if $i=j$, $2M_{i\bar{j}}=M_{i\bar{i}}+M_{j\bar{j}}$.
   Therefore, when $i\neq j$, we have
\begin{equation}
2M_{i\bar{j}} > M_{i\bar{i}}+M_{j\bar{j}}. \label{inequa1m}
\end{equation}

     Many authors argued recently that the slopes of Regge trajectories decrease with quark mass increase
 \cite{De-minLi2004,Ailin Zhang,Burakovsky-meson-relation,nonlinear,Kaidalov1982,residues-baryon,alpha-1991,Filipponi-slopesFAF,Basdevant1985}.
% the slope of the trajectories decreases with increasing quark mass in the mass region of the lowest excitations. % J.L. Basdevant and S. Boukraa, Z. Phys. C \textbf{28}, 413 (1985).
%  It is believed that the slope of the corresponding Regge trajectory for lighter hadrons should be larger than that for heavier hadrons.
  Therefore, $\frac{\alpha_{j\bar{j}}'}{\alpha_{i\bar{i}}'}<1$ when $j$-quark is heavier than $i$-quark.
Then, from Eq. (\ref{solution1}) one can have
\begin{equation}
\frac{1}{2M_{j\bar{j}}^2}\times[(4M_{i\bar{j}}^2-M_{i\bar{i}}^2-M_{j\bar{j}}^2)+\sqrt{(4M_{i\bar{j}}^2-M_{i\bar{i}}^2-M_{j\bar{j}}^2)^2-4M_{i\bar{i}}^2M_{j\bar{j}}^2} ] <1 .
\end{equation}
   From this relation, we obtain
\begin{equation}
\begin{cases}
\begin{aligned}
& 2M_{j\bar{j}}^2-(4M_{i\bar{j}}^2-M_{i\bar{i}}^2-M_{j\bar{j}}^2) > 0 ,   \\
&(4M_{i\bar{j}}^2-M_{i\bar{i}}^2-M_{j\bar{j}}^2)^2-4M_{i\bar{i}}^2M_{j\bar{j}}^2 < [2M_{j\bar{j}}^2-(4M_{i\bar{j}}^2-M_{i\bar{i}}^2-M_{j\bar{j}}^2)]^2 .
\end{aligned}
\end{cases}
\end{equation}
These two inequalities can be simplified to
%This relation can be simplified to
\begin{equation}
2M_{i\bar{j}}^2 < M_{i\bar{i}}^2+M_{j\bar{j}}^2. \label{inequa2m}
\end{equation}
  The relation (\ref{inequa2m}) can also be derived in the same way if we use the second equation
in Eq. (\ref{solution1}) considering $\frac{\alpha_{i\bar{j}}'}{\alpha_{i\bar{i}}'}<1$.

The baryon mass inequalities can be extracted in the same way as that in the meson case.
Then, we have
\begin{equation}
2M_{ijq} > M_{iiq}+M_{jjq}, \label{inequa1b}
\end{equation}
\begin{equation}
2M_{ijq}^2 < M_{iiq}^2+M_{jjq}^2.  \label{inequa2b}
\end{equation}

% Interestingly, Conducively, Helpfully, Usefully, Profitably, Effectively
%  It is very interesting that the linear mass inequalities and the quadratic mass inequalities we have observed are the concave and convex relations, respectively.
% One gives an upper limit, another gives a lower limit. For example, for the $M_{s\bar{s}}$, from relation (\ref{inequa1m}) and (\ref{inequa2m})
% They can give upper and lower mass limits for the mass of a hadron, respectively.

  It is very interesting that the inequalities (\ref{inequa1m}), (\ref{inequa2m}), (\ref{inequa1b}) and (\ref{inequa2b}) are the concave and convex relations.
  These mass inequalities can be used to give constrains (lower limits and upper limits) for masses of hadrons which have not been discovered.
For example, we have from the inequalities (\ref{inequa1m}) and (\ref{inequa2m}) that
\begin{equation}
\frac{M_{i\bar{i}}+M_{j\bar{j}}}{2} < M_{i\bar{j}} < \sqrt{\frac{M_{i\bar{i}}^2+M_{j\bar{j}}^2}{2}} \label{ineq-ms},
\end{equation}
in which one inequality gives an upper limit while the other gives a lower limit for $M_{i\bar{j}}$.
For baryons, we have from the inequalities (\ref{inequa1b}) and (\ref{inequa2b}) that
\begin{equation}
\frac{M_{iiq}+M_{jjq}}{2} < M_{ijq} < \sqrt{\frac{M_{iiq}^2+M_{jjq}^2}{2}} \label{ineq-bs}.
\end{equation}
   We will use Eqs. (\ref{ineq-ms}) and (\ref{ineq-bs}) to give mass ranges for mesons and baryons in Sec. III.

\subsection{Quadratic mass equalities }

  To evaluate the deviations of relations (\ref{inequa2m}) and (\ref{inequa2b}) from the equalities
that would be obtained by changing the signs of inequalities to equal signs,
we introduce a parameter $\delta$,
which is denoted by $\delta_{ij}^m$ for mesons,
\begin{equation}
\delta_{ij}^m=M_{i\bar{i}}^2+M_{j\bar{j}}^2-2M_{i\bar{j}}^2,  \label{delta-m}
\end{equation}
and by $\delta_{ij}^b$ for baryons,
\begin{equation}
\delta_{ij}^b=M_{iiq}^2+M_{jjq}^2-2M_{ijq}^2,
\end{equation}
where $i$, $j$ and $q$ are arbitrary light or heavy quarks.
     From relations (\ref{inequa2m}) and (\ref{inequa2b}), we know $\delta^{m(b)} >0$.
  It will be shown later that $\delta_{ij}^b$ is independent of $q$.
\\

   For mesons, from Eqs. (\ref{intercept1}) and (\ref{slope1a}), we have
\begin{equation}
a_{i\bar{i}}(0)-a_{i\bar{j}}(0)=a_{i\bar{j}}(0)-a_{j\bar{j}}(0) ,  \label{intercept-equal}
\end{equation}
\begin{equation}
\frac{1}{\alpha_{i\bar{i}}^\prime}-\frac{1}{\alpha_{i\overline{j}}^\prime}
=\frac{1}{\alpha_{i\overline{j}}^\prime}-\frac{1}{\alpha_{j\bar{j}}^\prime}.  \label{slope-equal}
\end{equation}
  Let
\begin{equation}
\lambda_{i} \equiv a_{n\bar{n}}(0)-a_{n\bar{i}}(0), \hspace{0.5cm}
\gamma_{i} \equiv \frac{1}{\alpha_{n\bar{i}}^\prime}-\frac{1}{\alpha_{n\bar{n}}^\prime},  \label{definition}
\end{equation}
where $n$ denotes light nonstrange quark $u$ or $d$.    % , and $A$ denotes $i$ or $j$
 Using Eqs. (\ref{intercept-equal}), (\ref{slope-equal}) and (\ref{definition}) we have
\begin{equation}
\lambda_{i} = a_{n\bar{n}}(0)-a_{n\bar{i}}(0) = a_{n\bar{i}}(0)-a_{i\bar{i}}(0) ,
\end{equation}
\begin{equation}
\gamma_{i} = \frac{1}{\alpha_{n\bar{i}}^\prime}-\frac{1}{\alpha_{n\bar{n}}^\prime}
    = \frac{1}{\alpha_{i\bar{i}}^\prime}-\frac{1}{\alpha_{n\bar{i}}^\prime} .
\end{equation}
  Hence,
\begin{equation}
a_{i\bar{i}}(0) = a_{n\bar{n}}(0)-2\lambda_{i} ,  \label{intercept-AA}
\end{equation}
\begin{equation}
 \frac{1}{\alpha_{i\bar{i}}^\prime} = \frac{1}{\alpha_{n\bar{n}}^\prime}+2\gamma_{i} . \label{slope-AA}
\end{equation}
   With the help of Eqs. (\ref{intercept-AA}) and (\ref{slope-AA}), we have from Eqs. (\ref{intercept1}) and (\ref{slope1a})
\begin{equation}
a_{i\bar{j}}(0)= \frac{1}{2} \left[ a_{i\bar{i}}(0)+a_{j\bar{j}}(0) \right]
           =a_{n\bar{n}}(0)-\lambda_i-\lambda_j \label{intercept-whole-m},
\end{equation}
\begin{equation}
\frac{1}{\alpha_{i\bar{j}}^\prime} =\frac{1}{2}\left( \frac{1}{\alpha_{i\bar{i}}^\prime}+\frac{1}{\alpha_{j\bar{j}}^\prime} \right)
          =\frac{1}{\alpha_{n\bar{n}}^\prime}+\gamma_i+\gamma_j \label{slope-whole-m}.
\end{equation}
   Similarly for baryons, from Eqs. (\ref{intercept2}) and (\ref{slope1b}), we have
\begin{equation}
a_{ijq}(0)=a_{nnn}(0)-\lambda_i-\lambda_j-\lambda_q ,
\label{intercept-whole}
\end{equation}
\begin{equation}
\frac{1}{\alpha_{ijq}^\prime}=\frac{1}{\alpha_{nnn}^\prime}+\gamma_i+\gamma_j+\gamma_q \label{slope-whole} ,
\end{equation}
where $\lambda_{x}\equiv a_{nnn}(0)-a_{nnx}(0)$, $\gamma_{x}\equiv\frac{1}{\alpha_{nnx}^\prime}-\frac{1}{\alpha_{nnn}^\prime}$ ($x$ denotes $i$, $j$ or $q$).
    It should be pointed out that the values of $\lambda_x$ and $\gamma_x$ can be different for different multiplets.

   For $n\bar{n}$ and $i\bar{j}$ states in a meson multiplet, from Eq. (\ref{regge1}), we have
\begin{equation}
J=a_{n\bar{n}}(0)+\alpha_{n\bar{n}}^\prime M_{n\bar{n}}^2, \label{transf-1}
\end{equation}
\begin{equation}
J=a_{i\bar{j}}(0)+\alpha_{i\bar{j}}^\prime M_{i\bar{j}}^2. \label{transf-2}
\end{equation}
   With the help of Eqs. (\ref{intercept-whole-m}), (\ref{slope-whole-m}) and (\ref{transf-1}), we have from Eq. (\ref{transf-2})
\begin{equation}
M_{i\bar{j}}^2=(\alpha_{n\bar{n}}^\prime M_{n\bar{n}}^2+\lambda_i+\lambda_j)(\frac{1}{\alpha_{n\bar{n}}'}+\gamma_{i}+\gamma_{j}). \label{transf-3}
\end{equation}
   Therefore, from Eqs. (\ref{delta-m}) and (\ref{transf-3}), we have
\begin{equation}
\begin{split}
\delta_{ij}^m =&(\alpha_{n\bar{n}}'M_{n\bar{n}}^2+2\lambda_{i})(\frac{1}{\alpha_{n\bar{n}}'}+2\gamma_{i})
                 +(\alpha_{n\bar{n}}'M_{n\bar{n}}^2+2\lambda_{j})(\frac{1}{\alpha_{n\bar{n}}'}+2\gamma_{j}) \\
               &-2(\alpha_{n\bar{n}}'M_{n\bar{n}}^2+\lambda_{i}+\lambda_{j})(\frac{1}{\alpha_{n\bar{n}}'}+\gamma_{i}+\gamma_{j}) \\
              =& 2(\lambda_i-\lambda_j)(\gamma_i-\gamma_j) .
\end{split}
\end{equation}

  For baryons, in the same way, we have
\begin{equation}
\begin{split}
\delta_{ij}^b =& M_{iiq}^2+M_{jjq}^2-2M_{ijq}^2\\
              =& (\alpha_{nnn}'M_{nnn}^2+2\lambda_{i}+\lambda_{q})(\frac{1}{\alpha_{nnn}'}+2\gamma_{i}+\gamma_{q})
                  +(\alpha_{nnn}'M_{nnn}^2+2\lambda_{j}+\lambda_{q})(\frac{1}{\alpha_{nnn}'}+2\gamma_{j}+\gamma_{q}) \\
               & -2(\alpha_{nnn}'M_{nnn}^2+\lambda_{i}+\lambda_{j}+\lambda_{q})(\frac{1}{\alpha_{nnn}'}+\gamma_{i}+\gamma_{j}+\gamma_{q}) \\
              =& 2(\lambda_i-\lambda_j)(\gamma_i-\gamma_j) \label{baryon-equal} .
\end{split}
\end{equation}
  It can be seen from Eq. (\ref{baryon-equal}) that $\delta_{ij}^b$ is independent of $q$.

   From Eq. (\ref{definition}), we know that $\lambda_n$=$\gamma_n$=$0$.
  Since we choose $m_i<m_j$, $\alpha_{ii}'>\alpha_{jj}'$.
 Hence from the definition of $\gamma_i$ (Eq. (\ref{definition})), we have $\gamma_i<\gamma_j$.
  Therefore, $0=\gamma_n<\gamma_s<\gamma_c<\gamma_b$.
   From Eqs. (\ref{solution1}) and (\ref{inequa1m}), we know that $\frac{\alpha_{jj}'}{\alpha_{ii}'}>\frac{M_{ii}}{M_{jj}}$.
 Hence ${\alpha_{jj}'}{M_{jj}}\cdot{M_{jj}} > {\alpha_{ii}'}{M_{ii}}\cdot{M_{ii}}$.
  With the help of Eqs. (\ref{regge1}) and (\ref{intercept-AA}), we have $\lambda_i<\lambda_j$.
   Therefore, $0=\lambda_n<\lambda_s<\lambda_c<\lambda_b$.
     Consequently, we have
$0< \delta_{ns}^m < \delta_{nc}^m < \delta_{nb}^m$, $0<
\delta_{cb}^m < \delta_{sb}^m < \delta_{nb}^m$, $0< \delta_{sc}^m
< \delta_{nc}^m$, and $0< \delta_{sc}^m < \delta_{sb}^m$.
    If we assume that $\gamma_s< \frac{1}{2}\gamma_c< \frac{1}{4}\gamma_b$ and $\lambda_s< \frac{1}{2}\lambda_c< \frac{1}{4}\lambda_b$, %(since $m_{s}\ll m_{c}\ll m_{b}$)
 with the above analysis, we can have $ \delta_{ns}^m < \delta_{sc}^m < \delta_{nc}^m < \delta_{cb}^m < \delta_{sb}^m < \delta_{nb}^m $.
    We will show later that these relations hold indeed.
    For baryons, we can have $ \delta_{ns}^b < \delta_{sc}^b < \delta_{nc}^b < \delta_{cb}^b < \delta_{sb}^b < \delta_{nb}^b $ in the same way.

  Inserting the corresponding masses into relation (\ref{delta-m}), we have the values of $\delta_{ij}^m$
for some meson multiplets which are shown in Table 2.
\begin{table}
Table 2.   The values of $\delta_{ij}^m$ for some multiplets (in units of GeV$^2$). \\
\begin{ruledtabular}

\begin{tabular}{c*{7}{l}}
                & $ \delta_{ns}^m$   & $\delta_{sc}^m$   & $\delta_{nc}^m$   & $\delta_{cb}^m$   & $\delta_{sb}^m$   & $\delta_{nb}^m$   \\\hline
1 $^{1}S_{0}$   &0.016      &1.623       &1.931     &16.898         &29.179       &30.769   \\
1 $^{3}S_{1}$   &0.015      &1.682       &2.125     &18.294         &31.930       &33.387   \\
1 $^{3}P_{2}$   &0.018         &1.785        &2.281     &18.042         &32.434       &34.018   \\
1 $^{1}P_{1}$   &         &             &2.198     &         &       &   \\
\end{tabular}

\end{ruledtabular}
\end{table}
       From Table 2, we can see that the relation $ \delta_{ns}^m < \delta_{sc}^m < \delta_{nc}^m < \delta_{cb}^m < \delta_{sb}^m < \delta_{nb}^m $ is indeed satisfied for different meson multiplets.
  These inequalities imply that the higher-order breaking effects become more pronounced with the quark mass increase. % rise with the quark mass increase.

\subsubsection{Mass relations for the $\frac{3}{2}^+$ multiplet}
  For the $\frac{3}{2}^+$ multiplet,
 noticing that $\delta_{ij}^{\frac{3}{2}^+}$ in the above relation (\ref{baryon-equal}) is independent of $q$,
 we have some equalities which are given in the following.

{%\footnotesize
\begin{subequations}
\label{eq:whole}
1) When $i=n$, $j=s$, $q=n,s,c,b$,
\begin{equation}
\delta_{ns}^{\frac{3}{2}^{+}}=M_{\Delta}^2+M_{\Xi^\ast}^2-2M_{\Sigma^\ast}^2=M_{\Sigma^\ast}^2+M_{\Omega}^2-2M_{\Xi^\ast}^2=M_{\Sigma_c^\ast}^2+M_{\Omega_c^\ast}^2-2M_{\Xi_c^\ast}^2=M_{\Sigma_b^\ast}^2+M_{\Omega_b^\ast}^2-2M_{\Xi_b^\ast}^2 \label{subeq:1} .
\end{equation}

2) When $i=n$, $j=c$, $q=n,s,c,b$,
\begin{equation}
\delta_{nc}^{\frac{3}{2}^{+}}=M_{\Delta}^2+M_{\Xi_{cc}^\ast}^2-2M_{\Sigma_c^\ast}^2=M_{\Sigma^\ast}^2+M_{\Omega_{cc}^\ast}^2-2M_{\Xi_c^\ast}^2=M_{\Sigma_c^\ast}^2+M_{\Omega_{ccc}}^2-2M_{\Xi_{cc}^\ast}^2=M_{\Sigma_b^\ast}^2+M_{\Omega_{bcc}^\ast}^2-2M_{\Xi_{bc}^\ast}^2 \label{subeq:2} .
\end{equation}

3) When $i=s$, $j=c$, $q=n,s,c,b$,
\begin{equation}
\delta_{sc}^{\frac{3}{2}^{+}}=M_{\Xi^\ast}^2+M_{\Xi_{cc}^\ast}^2-2M_{\Xi_c^\ast}^2=M_{\Omega}^2+M_{\Omega_{cc}^\ast}^2-2M_{\Omega_c^\ast}^2=M_{\Omega_c^\ast}^2+M_{\Omega_{ccc}}^2-2M_{\Omega_{cc}^\ast}^2=M_{\Omega_b^\ast}^2+M_{\Omega_{bcc}^\ast}^2-2M_{\Omega_{bc}^\ast}^2 \label{subeq:3} .
\end{equation}

4) When $i=n$, $j=b$, $q=n,s,c,b$,
\begin{equation}
\delta_{nb}^{\frac{3}{2}^{+}}=M_{\Delta}^2+M_{\Xi_{bb}^\ast}^2-2M_{\Sigma_b^\ast}^2=M_{\Sigma^\ast}^2+M_{\Omega_{bb}^\ast}^2-2M_{\Xi_b^\ast}^2=M_{\Sigma_c^\ast}^2+M_{\Omega_{bbc}^\ast}^2-2M_{\Xi_{bc}^\ast}^2=M_{\Sigma_b^\ast}^2+M_{\Omega_{bbb}}^2-2M_{\Xi_{bb}^\ast}^2 \label{subeq:4} .
\end{equation}

5) When $i=s$, $j=b$, $q=n,s,c,b$,
\begin{equation}
\delta_{sb}^{\frac{3}{2}^{+}}=M_{\Xi^\ast}^2+M_{\Xi_{bb}^\ast}^2-2M_{\Xi_b^\ast}^2=M_{\Omega}^2+M_{\Omega_{bb}^\ast}^2-2M_{\Omega_b^\ast}^2=M_{\Omega_c^\ast}^2+M_{\Omega_{bbc}^\ast}^2-2M_{\Omega_{bc}^\ast}^2=M_{\Omega_b^\ast}^2+M_{\Omega_{bbb}}^2-2M_{\Omega_{bb}^\ast}^2 \label{subeq:5} .
\end{equation}

6) When $i=c$, $j=b$, $q=n,s,c,b$,
\begin{equation}
\delta_{cb}^{\frac{3}{2}^{+}}=M_{\Xi_{cc}^\ast}^2+M_{\Xi_{bb}^\ast}^2-2M_{\Xi_{bc}^\ast}^2=M_{\Omega_{cc}^\ast}^2+M_{\Omega_{bb}^\ast}^2-2M_{\Omega_{bc}^\ast}^2=M_{\Omega_{ccc}}^2+M_{\Omega_{bbc}^\ast}^2-2M_{\Omega_{bcc}^\ast}^2=M_{\Omega_{bcc}^\ast}^2+M_{\Omega_{bbb}}^2-2M_{\Omega_{bbc}^\ast}^2  \label{subeq:6} .
\end{equation}

\end{subequations}
}

  From Eqs. (\ref{subeq:1})-(\ref{subeq:3}),
one can get the quadratic mass Eqs. (25)-(29) in Ref.
\cite{Burakovsky1997} derived by Burakovsky \emph{et al.}
   The linear forms of Eqs. (\ref{subeq:1})-(\ref{subeq:3}) were obtained
 by Hendry and Lichtenberg in the quark model \cite{Hendry-Lichtenberg-quarkmodel},
 by Verma and Khanna considering the second-order effects arising from the \underline{84} representation of SU(4) \cite{Verma-Khanna-SU4} and in the framework of $SU(8)$ symmetry \cite{Verma-Khanna-SU8},
 and by Singh \emph{et al}. studying SU(4) second-order mass-breaking effects with a dynamical consideration \cite{Singh-Verma-Khanna}
 (bottom baryons were not included in Refs. \cite{Burakovsky1997,Hendry-Lichtenberg-quarkmodel,Verma-Khanna-SU4,Verma-Khanna-SU8,Singh-Verma-Khanna}).
   The linear forms of Eqs. (\ref{subeq:1})-(\ref{subeq:6}) were derived
 by Singh and Khanna in the nonrelativistic additive quark model \cite{Singh-Khanna}
 and by Singh using broken SU(6) internal symmetry including second-order mass contributions \cite{Singh}.
    We will show some arguments in Sec. IV which support the quadratic form mass formulas for mesons and baryons
 rather than the linear form.

\subsubsection{Mass relations for the $\frac{1}{2}^+$ multiplet}
  For the $\frac{1}{2}^+$ multiplet, it is very different from the $\frac{3}{2}^+$ multiplet
because there are different ways for the spins of the constituent quarks to form the total spin $S=\frac{1}{2}$.
   Three constituent quarks in a $\frac{1}{2}^+$ baryon can be regarded as a quark and a scalar diquark or regarded as a quark and an axial-vector diquark.
  Regge slopes of $\Lambda$, $\Lambda_c$, $\Lambda_b$, $\Xi_{c}$ and $\Xi_{b}$ %, $\Xi_{bc}$, and $\Omega_{bc}$
are slightly bigger than those of $\Sigma$, $\Sigma_c$, $\Sigma_b$, $\Xi_{c}'$ and $\Xi_{b}'$, respectively,   % $\Xi_{bc}'$, and $\Omega_{bc}'$, respectively,
although sometimes they can be considered to be approximately equal \cite{Burakovsky1997}\cite{Kobylinsky-a}.
  Regge intercepts of $\Lambda$, $\Lambda_c$, $\Lambda_b$, $\Xi_{c}$ and $\Xi_{b}$ %, $\Xi_{bc}$, and $\Omega_{bc}$
are much bigger than those of $\Sigma$, $\Sigma_c$, $\Sigma_b$, $\Xi_{c}'$ and $\Xi_{b}'$,  respectively.  %$\Xi_{bc}'$, and $\Omega_{bc}'$, respectively.
    However, these cannot be reflected from Eqs. (\ref{intercept-whole}) and (\ref{slope-whole}).
    Therefore, some of the $\frac{1}{2}^+$ baryons may not be related as the $\frac{3}{2}^+$ baryons.
%   Therefore, not all the relations in Eq. (\ref{eq:whole}) for the $\frac{3}{2}^+$ multiplet have the corresponding form for the $\frac{1}{2}^+$ multiplet.
%   Thus, a model which can reflect these differences should be applied.  %     There is a study about these differences in instanton model \cite{Shuryad-Rosner-1989,Dey-Dey-Volkovitsky-1991}.
%     Then, we add a judgement that the linear form of the newly derived quadratic mass relations from Regge phenomenology for the $\frac{1}{2}^+$ multiplet must satisfy those relations (1)-(4) and (12)-(16) given in Ref. \cite{Dey-Dey-Volkovitsky-1991}.

    The `$Qqq'$' and `$QQ'q$' (where $q$ and $q'$ denote the light quarks while $Q$ and $Q'$ denote the heavy quarks $c$ or $b$) baryon states are believed to be described by the quark-diquark picture:
    Two light quarks $qq'$ are bound into a color antitriplet system with the size comparable to the QCD scale in the `$Qqq'$' baryon state \cite{Korner:1994nh,Guo-qq};
    Two heavy quarks $QQ'$ are bound into a small (compared with the QCD scale) color antitriplet system in the `$QQ'q$' baryon state \cite{Korner:1994nh,QQ-Falk-Doncheski-Guo-Tong}.
    The heavy baryons which are composed of a heavy quark and a light axial-vector diquark
($\Sigma_Q$, $\Xi_Q$ and $\Omega_Q$) belong to a $\textbf{6}$ representation of flavor $SU(3)$ \cite{PDG2006}.
    Therefore, $\delta_{ns}^{\frac{1}{2}^+}$ can be expressed as
\begin{equation}
\delta_{ns}^{\frac{1}{2}^+}=M_{\Sigma_c}^2+M_{\Omega_c}^2-2M_{\Xi_c'}^2=M_{\Sigma_b}^2+M_{\Omega_b}^2-2M_{\Xi_b'}^2. \label{delta-ns-12+}
\end{equation}
    For the doubly heavy baryons which are composed of a light quark and a heavy axial-vector diquark,
$\delta_{bc}^{\frac{1}{2}^+}$ can be expressed as
\begin{equation}
\delta_{bc}^{\frac{1}{2}^+}=M_{\Xi_{cc}}^2+M_{\Xi_{bb}}^2-2M_{\Xi_{bc}'}^2=M_{\Omega_{cc}}^2+M_{\Omega_{bb}}^2-2M_{\Omega_{bc}'}^2. \label{delta-bc-12+}
\end{equation}

   Since $\delta_{qq'}^{b}$ is determined by the dynamics of the light diquark system `$qq'$' inside a heavy baryon `$Qqq'$'
and since this dynamics is independent of flavor and spin of the
heavy quark due to the $SU(2)_f \bigotimes SU(2)_s$ symmetry in
the heavy quark limit \cite{HQET},
  we assume that $\delta_{ns}^{\frac{1}{2}^{+}}$ for the $\frac{1}{2}^+$ charmed (bottom) sextet
equals $\delta_{ns}^{\frac{3}{2}^{+}}$ for the $\frac{3}{2}^+$
charmed (bottom) sextet,
$\delta_{ns}^{\frac{1}{2}^{+}}=\delta_{ns}^{\frac{3}{2}^{+}}$.
% because both the light diquark systems in sextets in two multiplets have spin S=1 (vector system).
   This relation holds exactly when the masses of charmed and bottom quarks are taken to be infinitely large.
   Deviations from this relation are due to $\frac{1}{m_c}$ and $\frac{1}{m_b}$ corrections.
  Then, one can have
\begin{equation}
M_{\Sigma_c}^2+M_{\Omega_c}^2-2M_{\Xi_c'}^2=M_{\Sigma_c^\ast}^2+M_{\Omega_c^\ast}^2-2M_{\Xi_c^\ast}^2,
\end{equation}
\begin{equation}
M_{\Sigma_b}^2+M_{\Omega_b}^2-2M_{\Xi_b'}^2=M_{\Sigma_b^\ast}^2+M_{\Omega_b^\ast}^2-2M_{\Xi_b^\ast}^2.
\end{equation}
There are two linear mass equations similar to the above quadratic
mass equations,
\begin{equation}
M_{\Sigma_c}+M_{\Omega_c}-2M_{\Xi_c'}=M_{\Sigma_c^\ast}+M_{\Omega_c^\ast}-2M_{\Xi_c^\ast},
\end{equation}
\begin{equation}
M_{\Sigma_b}+M_{\Omega_b}-2M_{\Xi_b'}=M_{\Sigma_b^\ast}+M_{\Omega_b^\ast}-2M_{\Xi_b^\ast},
\end{equation}
which were extracted by Jenkins in the $1/m_Q$ and $1/N_c$
expansions \cite{sextet-Jenkins}.
% Incidentally the linear form of the above two relations is extracted by Jenkins in the $1/m_Q$ and $1/N_c$ expansions \cite{sextet-Jenkins}.
%
Similarly, assuming that
$\delta_{bc}^{\frac{1}{2}^{+}}$=$\delta_{bc}^{\frac{3}{2}^{+}}$,
one can have
\begin{equation}
M_{\Xi_{cc}}^2+M_{\Xi_{bb}}^2-2M_{\Xi_{bc}'}^2=M_{\Omega_{cc}}^2+M_{\Omega_{bb}}^2-2M_{\Omega_{bc}'}^2=M_{\Xi_{cc}^\ast}^2+M_{\Xi_{bb}^\ast}^2-2M_{\Xi_{bc}^\ast}^2=M_{\Omega_{cc}^\ast}^2+M_{\Omega_{bb}^\ast}^2-2M_{\Omega_{bc}^\ast}^2.
\end{equation}

%  With these relations, we still can not relate some masses of the baryon states in the $\frac{1}{2}^+$ multiplet % in which the diquark system have spin $S_{d}=0$ (scalar system).
%  We will study the $\frac{1}{2}^+$ baryon multiplet with the help of the Bethe-Salpeter equation in quark-diquark picture later.

   From Eq. (\ref{eq:whole}), we can have a relation for the $\frac{3}{2}^+$ baryons,
\begin{equation}
(M_{\Omega_{cc}^*}^2-M_{\Xi_{cc}^*}^2)+(M_{\Xi^*}^2-M_{\Sigma^*}^2)=(M_{\Omega_c^*}^2-M_{\Sigma_c^*}^2).  \label{Omega-cc}
\end{equation}
   %There is another relation for the $\frac{1}{2}^+$ multiplet identical to the corresponding relation for the $\frac{3}{2}^+$ multiplet which can be obtained from Regge phenomenology
   Its corresponding relation for the $\frac{1}{2}^+$ baryons is
\begin{equation}
(M_{\Omega_{cc}}^2-M_{\Xi_{cc}}^2)+(M_\Xi^2-M_\Sigma^2)=(M_{\Omega_c}^2-M_{\Sigma_c}^2) \label{Omega-cc-1} .
\end{equation}
   The linear form of Eq.(\ref{Omega-cc-1}) can satisfy the instanton model \cite{instanton-model}
and has been given by Verma and Khanna considering the second-order effects arising from the $\underline{84}$ representation of $SU(4)$ \cite{Verma-Khanna-SU4}.
%   Eq. (31) in Ref. \cite{Burakovsky1997},
    A different relation,
\begin{equation}
(M_{\Omega_{cc}}^2-M_{\Xi_{cc}}^2)+\left( \frac{3M_\Lambda^2+M_\Sigma^2}{4}-M_N^2 \right)=2\left( \frac{M_{\Xi_c}^2+M_{\Xi_c'}^2}{2}-\frac{3M_{\Lambda_c}^2+M_{\Sigma_c}^2}{4} \right) \label{Omega-cc-2} ,
\end{equation}
has been proposed in Ref. \cite{Burakovsky1997}.
   However, the linear form of Eq. (\ref{Omega-cc-2}) cannot satisfy the instanton model \cite{instanton-model}.
   Furthermore, the value of $(M_{\Omega_{cc}}^2-M_{\Xi_{cc}}^2)$ given by Eq. (\ref{Omega-cc-1}) ($\sim$ 0.94 $GeV^2$)
 is close to the value of $(M_{\Omega_{cc}^*}^2-M_{\Xi_{cc}^*}^2)$ given by Eq. (\ref{Omega-cc}) ($\sim$ 0.89 $GeV^2$)
 while the value of $(M_{\Omega_{cc}}^2-M_{\Xi_{cc}}^2)$ given by Eq. (\ref{Omega-cc-2}) ($\sim$ 1.39 $GeV^2$)
 is much larger.
%   Furthermore, the value of $(M_{\Omega_{cc}}-M_{\Xi_{cc}})$ given by Eq. (\ref{Omega-cc-2}) (about 200 $MeV$) is larger than the usual predictions (about 120Mev).
   We will use Eq. (\ref{Omega-cc-1}) rather than Eq. (\ref{Omega-cc-2}) to extract the mass of $\Omega_{cc}$ in Sec. III.

\section{Some applications}

    In this section, we will apply the relations we have obtained in Sec. II to discuss
 the mass ranges of mesons and baryons, the masses of the $\bar{b}c$ and $s\bar{s}$ meson states,
the properties of $\Xi_{cc}^+(3520)$, the parameters of the Regge trajectories for the $\frac{1}{2}^+$ and $\frac{3}{2}^+$ multiplet,
and the properties of the charm-strange baryons (some of which have just been observed).

\subsection{Mass ranges of mesons and baryons}

% Application of inequalities
     Using Eqs. (\ref{ineq-ms}) and (\ref{ineq-bs}), we calculate the upper and lower mass limits for
 some meson states ($s\bar{s}$, $c\bar{n}$, $\bar{b}n$, $\bar{b}c$, $c\bar{s}$ and $\bar{b}s$) of different multiplets
 and some baryon states of $\frac{1}{2}^+$ and $\frac{3}{2}^+$ multiplets.
  The results for mesons are shown in Table 3-1 and Table 3-2 in comparison with the measured meson masses \cite{PDG2006}.
  The results for baryons are shown in Table 4 in comparison with the measured baryon masses \cite{PDG2006}.

\begin{table}%[H]
    Table 3-1.   The numerical results for upper and lower limits for the masses of mesons ($s\bar{s}$, $c\bar{n}$ and $\bar{b}n$)
 obtained from Eqs. (\ref{inequa1m}) and (\ref{inequa2m}) in comparison with the experimental data (in units of GeV).
\begin{ruledtabular}

\begin{tabular}{lcc}
 $\mathcal{N}^{2S+1}L_{J}$    &Inequalities                                                                       &Lower and upper limits  \\\hline

$s\bar{s}$ sector   &$\sqrt{2M_{n\bar{s}}^2-M_{n\bar{n}}^2}<M_{s\bar{s}}<2M_{n\bar{s}}-M_{n\bar{n}}$    &   \\

1 $^{1}S_{0}$       &$\sqrt{2M_{K}^2-M_{\pi}^2}<M_{s\bar{s}}<2M_{K}-M_{\pi}$                            & $0.687<M_{s\bar{s}}<0.854$ \\
1 $^{3}S_{1}$       &$\sqrt{2M_{K^*}^2-M_{\rho}^2}<M_{s\bar{s}}<2M_{K^*}-M_{\rho}$                      & $0.998<M_{s\bar{s}}<1.012$ \\
1 $^{3}P_{2}$       &$\sqrt{2M_{K_2^*}^2-M_{a_2(1320)}^2}<M_{s\bar{s}}<2M_{K_2^*}-M_{a_2(1320)}$        & $1.538<M_{s\bar{s}}<1.547$ \\
1 $^{1}D_{2}$       &$\sqrt{2M_{K_2(1770)}^2-M_{\pi_2(1670)}^2}<M_{s\bar{s}}<2M_{K_2(1770)}-M_{\pi_2(1670)}$        & $1.868<M_{s\bar{s}}<1.874$ \\
1 $^{3}D_{3}$       &$\sqrt{2M_{K_3^*}^2-M_{\rho_3}^2}<M_{s\bar{s}}<2M_{K_3^*}-M_{\rho_3}$              & $1.859<M_{s\bar{s}}<1.863$ \\

\\
$c\bar{n}$ sector   &$(M_{n\bar{n}}+M_{c\bar{c}})/2<M_{c\bar{n}}<\sqrt{(M_{n\bar{n}}^2+M_{c\bar{c}}^2)/2}$      &  \\

1 $^{1}S_{0}$       &$(M_\pi+M_{\eta_c(1S)})/2<M_{D}<\sqrt{(M_\pi^2+M_{\eta_c(1S)}^2)/2}$               & $1.559<1.867(exp.)<2.110$ \\
1 $^{3}S_{1}$       &$(M_\rho+M_{J/\psi(1S)})/2<M_{D^*}<\sqrt{(M_\rho^2+M_{J/\psi(1S)}^2)/2}$           & $1.936<2.008(exp.)<2.257$  \\
1 $^{3}P_{2}$       &$(M_{a_2(1320)}+M_{\chi_{c2}(1P)})/2<M_{D_2^*}<\sqrt{(M_{a_2(1320)}^2+M_{\chi_{c2}(1P)}^2)/2}$     & $2.437<2.460(exp.)<2.682$ \\
1 $^{1}P_{1}$       &$(M_{b_1(1235)}+M_{h_c(1P)})/2<M_{D_1(2420)}<\sqrt{(M_{b_1(1235)}^2+M_{h_c(1P)}^2)/2}$             & $2.378<2.423(exp.)<2.640$ \\
1 $^{3}P_{1}$       &$(M_{a_1(1260)}+M_{\chi_{c1}(1P)})/2<M_{D_1(1^3P_1)}<\sqrt{(M_{a_1(1260)}^2+M_{\chi_{c1}(1P)}^2)/2}$     & $2.370<M_{D_1(1^3P_1)}<2.630$ \\
1 $^{3}D_{1}$       &$(M_{\rho(1700)}+M_{\psi(3770)})/2<M_{D^*(1^{3}D_{1})}<\sqrt{(M_{\rho(1700)}^2+M_{\psi(3770)}^2)/2}$  & $2.746<M_{D^*(1 ^{3}D_{1})}<2.931$ \\
2 $^{1}S_{0}$       &$(M_{\pi(1300)}+M_{\eta_c(2S)})/2<M_{D(2^{1}S_{0})}<\sqrt{(M_{\pi(1300)}^2+M_{\eta_c(2S)}^2)/2}$   & $2.419<M_{D(2^{1}S_{0})}<2.756$ \\
2 $^{3}S_{1}$       &$(M_{\rho(1450)}+M_{\psi(2S)})/2<M_{D^*(2^{3}S_{1})}<\sqrt{(M_{\rho(1450)}^2+M_{\psi(2S)}^2)/2}$   & $2.573<M_{D^*(2^{3}S_{1})}<2.803$ \\

\\
$\bar{b}n$ sector   &$(M_{n\bar{n}}+M_{b\bar{b}})/2<M_{\bar{b}n}<\sqrt{(M_{n\bar{n}}^2+M_{b\bar{b}}^2)/2}$      &  \\

1 $^1S_0$       &$(M_\pi+M_{\eta_b(1S)})/2<M_{B}<\sqrt{(M_\pi^2+M_{\eta_b(1S)}^2)/2}$                       & $4.719<5.279(exp.)<6.577$ \\
1 $^3S_1$       &$(M_\rho+M_{\Upsilon(1S)})/2<M_{B^*}<\sqrt{(M_\rho^2+M_{\Upsilon(1S)}^2)/2}$               & $5.118<5.325(exp.)<6.712$  \\
1 $^3P_2$       &$(M_{a_2(1320)}+M_{\chi_{b2}(1P)})/2<M_{B_2^*}<\sqrt{(M_{a_2(1320)}^2+M_{\chi_{b2}(1P)}^2)/2}$        & $5.615<5.743(exp.)<7.071$ \\
1 $^3P_1$       &$(M_{a_1(1260)}+M_{\chi_{b1}(1P)})/2<M_{B_1 (1^3P_1)}<\sqrt{(M_{a_1(1260)}^2+M_{\chi_{b1}(1P)}^2)/2}$     & $5.561<M_{B_1 (1^3P_1)}<7.049$ \\
2 $^3S_1$       &$(M_{\rho(1450)}+M_{\Upsilon(2S)})/2<M_{B^*(2^3S_1)}<\sqrt{(M_{\rho(1450)}^2+M_{\Upsilon(2S)}^2)/2}$       & $5.741<M_{B^*(2^3S_1)}<7.162$  \\
2 $^3P_2$       &$(M_{a_2(1700)}+M_{\chi_{b2}(2P)})/2<M_{B_2^*(2^3P_2)}<\sqrt{(M_{a_2(1700)}^2+M_{\chi_{c2}(2P)}^2)/2}$     & $6.000<M_{B_2^*(2^3P_2)}<7.363$ \\
\end{tabular}

\end{ruledtabular}
\end{table}

\begin{table}%[H]
   Table 3-2.   The numerical results for upper and lower limits for the masses of mesons ($\bar{b}c$, $c\bar{s}$ and $\bar{b}s$)
 obtained from Eqs. (\ref{inequa1m}) and (\ref{inequa2m}) in comparison with the experimental data (in units of GeV).
\begin{ruledtabular}

\begin{tabular}{lcc}
 $\mathcal{N}^{2S+1}L_{J}$    &Ineqalities                                                                        &Lower and upper limits  \\\hline

$\bar{b}c$ sector   &$(M_{c\bar{c}}+M_{b\bar{b}})/2<M_{\bar{b}c}<\sqrt{(M_{c\bar{c}}^2+M_{b\bar{b}}^2)/2}$      &  \\

1 $^1S_0$       &$(M_{\eta_c(1S)}+M_{\eta_b(1S)})/2<M_{B_c}<\sqrt{(M_{\eta_c(1S)}^2+M_{\eta_b(1S)}^2)/2}$               & $6.140<6.286(exp.)<6.906$ \\
1 $^3S_1$       &$(M_{J/\psi(1S)}+M_{\Upsilon(1S)})/2<M_{B_c^*}<\sqrt{(M_{J/\psi(1S)}^2+M_{\Upsilon(1S)}^2)/2}$       & $6.279<M_{B_c^*}<7.039$  \\
1 $^3P_2$       &$(M_{\chi_{c2}(1P)}+M_{\chi_{b2}(1P)})/2<M_{B_{c2}^*}<\sqrt{(M_{\chi_{c2}(1P)}^2+M_{\chi_{b2}(1P)}^2)/2}$     & $6.734<M_{B_{c2}^*}<7.446$ \\
1 $^3P_0$       &$(M_{\chi_{c0}(1P)}+M_{\chi_{b0}(1P)})/2<M_{B_{c0}^*}<\sqrt{(M_{\chi_{c0}(1P)}^2+M_{\chi_{b0}(1P)}^2)/2}$     & $6.637<M_{B_{c0}^*}<7.378$ \\
1 $^3P_1$       &$(M_{\chi_{c1}(1P)}+M_{\chi_{b1}(1P)})/2<M_{B_{c1}(1^3P_1)}<\sqrt{(M_{\chi_{c1}(1P)}^2+M_{\chi_{b1}(1P)}^2)/2}$     & $6.702<M_{B_{c1}(1^3P_1)}<7.423$ \\
2 $^3S_1$       &$(M_{\psi(2S)}+M_{\Upsilon(2S)})/2<M_{B_c^*(2^3S_1)}<\sqrt{(M_{\psi(2S)}^2+M_{\Upsilon(2S)}^2)/2}$       & $6.855<M_{B_c^*(2^3S_1)}<7.552$  \\
\\
$c\bar{s}$ sector   &$(M_{c\bar{c}}+M_{s\bar{s}})/2<M_{c\bar{s}}<\sqrt{(M_{c\bar{c}}^2+M_{s\bar{s}}^2)/2}$      &  \\

1 $^1S_0$       &$(M_{\eta_c(1S)}+M_{s\bar{s}(1^1S_0)})/2<M_{D_s}<\sqrt{(M_{\eta_c(1S)}^2+M_{s\bar{s}(1^1S_0)}^2)/2}$               & $1.834<1.968(exp.)<2.163$ \\
1 $^3S_1$       &$(M_{J/\psi(1S)}+M_{s\bar{s}(1^3S_1)})/2<M_{D_s^*}<\sqrt{(M_{J/\psi(1S)}^2+M_{s\bar{s}(1^3S_1)}^2)/2}$             & $2.058<2.112(exp.)<2.305$  \\
1 $^3P_2$       &$(M_{\chi_{c2}(1P)}+M_{s\bar{s}(1^3P_2)})/2<M_{D_{s2}^*}<\sqrt{(M_{\chi_{c2}(1P)}^2+M_{s\bar{s}(1^3P_2)}^2)/2}$    & $2.541<2.574(exp.)<2.736$ \\
\\
$\bar{b}s$ sector   &$(M_{b\bar{b}}+M_{s\bar{s}})/2<M_{\bar{b}s}<\sqrt{(M_{b\bar{b}}^2+M_{s\bar{s}}^2)/2}$      &  \\

1 $^1S_0$       &$(M_{\eta_b(1S)}+M_{s\bar{s}(1^1S_0)})/2<M_{B_s}<\sqrt{(M_{\eta_b(1S)}^2+M_{s\bar{s}(1^1S_0)}^2)/2}$               & $4.994<5.368(exp.)<6.594$ \\
1 $^3S_1$       &$(M_{\Upsilon(1S)}+M_{s\bar{s}(1^3S_1)})/2<M_{B_s^*}<\sqrt{(M_{\Upsilon(1S)}^2+M_{s\bar{s}(1^3S_1)}^2)/2}$       & $5.240<5.413(exp.)<6.728$  \\
1 $^3P_2$       &$(M_{\chi_{b2}(1P)}+M_{s\bar{s}(1^3P_2)})/2<M_{B_{s2}^*}<\sqrt{(M_{\chi_{b2}(1P)}^2+M_{s\bar{s}(1^3P_2)}^2)/2}$     & $5.719<5.840(exp.)<7.091$ \\
\end{tabular}

\end{ruledtabular}
\end{table}

\begin{table}%[H]
Table 4.   The numerical results for upper and lower limits for the masses of baryons obtained
 from Eqs. (\ref{inequa1b}) and (\ref{inequa2b}) in comparison with the experimental data (in units of GeV).
\begin{ruledtabular}

\begin{tabular}{lc}

            $J^P=\frac{1}{2}^+$ inequalities                        &Lower and upper limits  \\\hline

$(M_N+M_{\Xi})/2<(3M_{\Lambda}+M_{\Sigma})/4<\sqrt{(M_N^2+M_{\Xi}^2)/2}$                & $1.128<1.135(exp.)<1.144$ \\
$(M_{\Sigma_c}+M_{\Omega_c})/2<M_{\Xi_c}<\sqrt{(M_{\Sigma_c}^2+M_{\Omega_c}^2)/2}$      & $2.576<2.577(exp.)<2.578$ \\
$(M_N+M_{\Xi_{cc}})/2<(3M_{\Lambda_c}+M_{\Sigma_c})/4<\sqrt{(M_N^2+M_{\Xi_{cc}}^2)/2}$  & $3.156<M_{\Xi_{cc}}<3.718$ \\
$(M_N+M_{\Xi_{bb}})/2<(3M_{\Lambda_b}+M_{\Sigma_b})/4<\sqrt{(M_N^2+M_{\Xi_{bb}}^2})/2$  & $7.965<M_{\Xi_{bb}}<10.403$ \\ \\\hline

            $J^P=\frac{3}{2}^+$ inequalities                        &Lower and upper limits   \\\hline

$(M_{\Delta}+M_{\Xi^*})/2<M_{\Sigma^*}<\sqrt{(M_{\Delta}^2+M_{\Xi^*}^2)/2}$             & $1.383<1.385(exp.)<1.391$ \\
$(M_{\Sigma^*}+M_\Omega)/2<M_{\Xi^*}<\sqrt{(M_{\Sigma^*}^2+M_\Omega^2)/2}$              & $1.529<1.533(exp.)<1.535$ \\

$\sqrt{2M_{\Xi_c^*}^2-M_{\Sigma_c^*}^2}<M_{\Omega_c^*}<2M_{\Xi_c^*}-M_{\Sigma_c^*}$     & $2.766<2.768(exp.)<2.778$ \\

$\sqrt{2M_{\Sigma_c^*}^2-M_{\Delta}^2}<M_{\Xi_{cc}^*}<2M_{\Sigma_c^*}-M_{\Delta}$     & $3.341<M_{\Xi_{cc}^*}<3.804$ \\
$\sqrt{2M_{\Xi_c^*}^2-M_{\Xi}^2}<M_{\Xi_{cc}^*}<2M_{\Xi_c^*}-M_{\Xi}$                 & $3.414<M_{\Xi_{cc}^*}<3.759$ \\

$\sqrt{2M_{\Xi_c^*}^2-M_{\Sigma}^2}<M_{\Omega_{cc}^*}<2M_{\Xi_c^*}-M_{\Sigma}$  & $3.477<M_{\Omega_{cc}^*}<3.908$ \\
$\sqrt{2M_{\Omega_c}^2-M_\Omega^2}<M_{\Omega_{cc}^*}<2M_{\Omega_c}-M_\Omega$ & $3.544<M_{\Omega_{cc}^*}<3.869$ \\

$\sqrt{2M_{\Sigma_b^*}^2-M_{\Delta}^2}<M_{\Xi_{bb}^*}<2M_{\Sigma_b^*}-M_{\Delta}$     & $8.156<M_{\Xi_{bb}^*}<10.433$ \\
\end{tabular}

\end{ruledtabular}
\end{table}

   The masses of the pure $s\bar{s}$ states cannot be directly measured experimentally
because of the usual mixing of the pure isoscalar $n\bar{n}$ and
$s\bar{s}$ states.
   The way to extract masses of the pure $s\bar{s}$ states will be displayed in the next section. % (High-power mass equalities).
   In calculating the mass limits about the $c\bar{s}$ and $\bar{b}s$ states in Table 3-2,
 we approximately use the values of $\sqrt{2M_K^2-M_\pi^2}$ (given by the quadratic GMO formula $M_\pi^2+M_{s \bar{s}(1 ^1S_0)}^2=2M_K^2$), $M_\phi$ and $M_{f_2'(1525)}$ to replace
$M_{s \bar{s}(1 ^1S_0)}$, $M_{s\bar{s}(1 ^3S_1)}$ and
$M_{s\bar{s}(1 ^3P_2)}$, respectively. $f_2'(1525)$ was proved to
be a nearly pure tensor $s\bar{s}$ state ($\sim$ 98.2 $\%$)
\cite{f2'(1525)-ss}.
   These approximations shift the mass limits of the $c\bar{s}$ and $\bar{b}s$ states only a few MeV.

   It can be seen from Tables 3-1, 3-2 and 4 that the inequalities (\ref{ineq-ms}) and (\ref{ineq-bs})
 (which were given from the inequalities (\ref{inequa1m}), (\ref{inequa2m}), (\ref{inequa1b}) and (\ref{inequa2b}))
 agree well with the existing experimental data \cite{PDG2006}.
   The inequalities (\ref{ineq-ms}) and (\ref{ineq-bs}) also
give predictions for the mass ranges of some hadrons which have not been observed.
%
%   The mass ranges for the pure $s\bar{s}$ states are very narrow, for charmed hadrons wide and for bottom hadrons large.
%   These feature manifest that both the linear and the quadratic GMO formulas
% can not be directly generalized to the charmed and bottom hadron.
%   Some of these mass ranges may be helpful for the discovery of the unobserved hadron states and the spin-parity assignment of these states.
%   The mass ranges involving bottom hadrons are so large that they are not useful for the spin-parity assignment of these states,
% but they may be helpful to reconstruct the mass window for the detector when searching these unobserved states.
%
    More detailed discussions about the inequalities derived in this work and those in Refs. \cite{inequalities-history, inequalities-QCD1,inequalities-QCD2,inequalities-QCD3}
 will be given in Sec. IV.
% These constraints may be useful for the discovery of the unobserved baryon states and the assignment of the new observed baryon states.

\subsection{Masses of the $\bar{b}c$ and $s\bar{s}$ meson states}

\subsubsection{Masses of the $\bar{b}c$ meson states}
   The $\bar{b}c$ (or $b\bar{c}$) meson states are special systems with two heavy quarks of different flavors.
 The presence of both such quarks impacts on the production, decay and mass properties of the $\bar{b}c$ mesons.
 Until recently, only the pseudoscalar mesons $B_c^{\pm}$ have been observed experimentally \cite{PDG2006, com-bc-exp-2,Bc-new-CDF}.
 The copious productions of $B_c$ mesons and their radial and orbital excitations are expected
at the experimental facilities such as the Large Hadron Collider
(LHC) at CERN.
  The masses of $\bar{b}c$ mesons have been predicted in many different approaches \cite{De-minLi2004,comp-bc-1,comp-bc-2,comp-bc-3,comp-bc-4,comp-bc-5,comp-bc-6,comp-bc-7,comp-bc-8,comp-bc-9,comp-bc-10,Baldicchi:2000-bc-11,comp-bc-12}.
% There are many theoretical studies on this project such as lattice QCD calculations [ ] and potential models [ ].
    In the following, we will use Eq. (\ref{high-equal-m}) to calculate the masses of $B_c$, $B_c^*$ and $B_{c2}^*$ meson states
and compare the results with those given in Refs.
\cite{De-minLi2004,comp-bc-1,comp-bc-2,comp-bc-3,comp-bc-4,comp-bc-5,comp-bc-6,comp-bc-7,comp-bc-8,comp-bc-9,comp-bc-10,Baldicchi:2000-bc-11,comp-bc-12}.

  For the $1 ^{1}S_{0}$ multiplet, when $i=n$, $j=c$, and $k=b$, inserting the masses of $\pi$, $\eta_c(1S)$, $\eta_b(1S)$, $D$ and $B$
into Eq. (\ref{high-equal-m}), the mass of $B_c$ can be extracted.
  For the $1 ^{3}S_{1}$ multiplet, when $i=n$, $j=c$, and $k=b$, inserting the masses of $\rho$, $J/\psi(1S)$, $\Upsilon(1S)$, $D^*$ and $B^*$
into Eq. (\ref{high-equal-m}), the mass of $B_c^*$ can be
extracted.
  For the $1 ^{3}P_{2}$ multiplet, when $i=n$, $j=c$, and $k=b$, inserting the masses of $a_2(1320)$, $\chi_{c2}(1P)$, $\chi_{b2}(1P)$, $D_2^*(2460)$
and $B_2^*(5740)$ which was observed recently \cite{B2-D0-CDF}
into Eq. (\ref{high-equal-m}), the mass of $B_{c2}^*$ can be
extracted.
  Comparison of the masses of $B_c$, $B_c^*$ and $B_{c2}^*$ extracted in the present work and those given by other references is shown in Table 5.
  The application of Eq. (\ref{equa-double}) (baryon case) will be performed in Subsection D of this section.

  If Eq. (\ref{intercept1}) (the additivity of inverse slopes) were replaced by Eq. (\ref{factorization-slope-m}) (the factorization of slopes)
in the derivation of Eq. (\ref{high-equal-m}), we would have the
following equation instead of Eq. (\ref{high-equal-m}),
\begin{equation}
\begin{split}
&\frac{(2M_{i\bar{j}}^4-M_{i\bar{i}}^2 M_{j\bar{j}}^2) + 2M_{i\bar{j}}^2 \sqrt{(2M_{i\bar{j}}^4-M_{i\bar{i}}^2 M_{j\bar{j}}^2)}}{M_{j\bar{j}}^4}
 \\ =& \frac{[(2M_{i\bar{k}}^4-M_{i\bar{i}}^2 M_{k\bar{k}}^2) +2 M_{i\bar{k}}^2 \sqrt{(2M_{i\bar{k}}^4-M_{i\bar{i}}^2 M_{k\bar{k}}^2)}]/{M_{k\bar{k}}^4}}
{[(2M_{j\bar{k}}^4-M_{j\bar{j}}^2M_{k\bar{k}}^2) + 2M_{j\bar{k}}^2 \sqrt{(2M_{j\bar{k}}^4-M_{j\bar{j}}^2 M_{k\bar{k}}^2)}] / {M_{k\bar{k}}^4}}  \label{high-equal-factor}.
\end{split}
\end{equation}
 Applying this equation to the $1 ^{1}S_{0}$, $1 ^{3}S_{1}$ and $1 ^{3}P_{2}$ multiplets
we would extract the masses of $B_c$, $B_c^*$ and $B_{c2}^*$
which are also shown in Table 5.

     In Ref. \cite{Burakovsky-meson-relation}, under the approximation that mesons in the light quark sector have the common Regge slopes,
 a 14th power meson mass relation,
\begin{equation}
\begin{split}
& [(M_{s\bar{s}}^2-M_{n\bar{n}}^2)(M_{c\bar{c}}^2 M_{n\bar{b}}^2
(M_{c\bar{s}}^2-M_{c\bar{n}}^2) + M_{b\bar{b}}^2 M_{c\bar{n}}^2
(M_{s\bar{b}}^2-M_{n\bar{b}}^2) ) - M_{n\bar{n}}^2
(M_{c\bar{c}}^2+M_{b\bar{b}}^2) (M_{c\bar{s}}^2-M_{c\bar{n}}^2)
(M_{s\bar{b}}^2-M_{n\bar{b}}^2) ]
\\ & \times [(M_{s\bar{s}}^2-M_{n\bar{n}}^2)(M_{n\bar{b}}^2 (M_{c\bar{s}}^2-M_{c\bar{n}}^2) + M_{c\bar{n}}^2 (M_{s\bar{b}}^2-M_{n\bar{b}}^2) )-2M_{n\bar{n}}^2 (M_{c\bar{s}}^2-M_{c\bar{n}}^2) (M_{s\bar{b}}^2-M_{n\bar{b}}^2) ]
\\ =& 4M_{b\bar{c}}^2  (M_{c\bar{s}}^2-M_{c\bar{n}}^2) (M_{s\bar{b}}^2-M_{n\bar{b}}^2)  (M_{c\bar{n}}^2M_{s\bar{s}}^2-M_{c\bar{s}}^2M_{n\bar{n}}^2) (M_{n\bar{b}}^2M_{s\bar{s}}^2-M_{s\bar{b}}^2M_{n\bar{n}}^2) \label{14th-power},
\end{split}
\end{equation}
was derived to predict the mass of $B_c^*$ with the value $M_{B_c^*}$=6.285 GeV.
  The results of applying Eq. (\ref{14th-power}) with the existing experimental data \cite{PDG2006} for the $1 ^{1}S_{0}$, $1 ^{3}S_{1}$ and $1 ^{3}P_{2}$ multiplets
to extract the masses of $B_c$, $B_c^*$ and $B_{c2}^*$ are also
shown in Table 5.

\begin{table}
Table 5.   The masses of $B_{c}$, $B_c^*$, and $B_{c2}^*$ (in units of GeV). \\
\begin{ruledtabular}

\begin{tabular}{c c *{8}{l}}
States ($\mathcal{N} ^{2S+1}L_{J}$) &Present work       &Eq. (\ref{high-equal-factor})  &Eq. (\ref{14th-power})        &Exp.                 &\cite{De-minLi2004}    &\cite{comp-bc-3}        &\cite{comp-bc-1}      &\cite{comp-bc-4}   &\cite{comp-bc-5}     \\\hline
$B_{c}$ (1 $^{1}S_{0}$)             &6.264            &6.404                      &6.142                            &6.276\footnotemark[1]   &6.263                 &6.270                  &6.253                  &6.264              &6.247            \\
$B_c^*$ (1 $^{3}S_{1}$)             &6.356           &6.502                      &6.292                             &                        &6.354                 &6.332                  &6.317                  &6.337              &6.308           \\
$B_{c2}^*$  (1 $^3P_2$)             &6.814            &6.940                      &6.767                            &                         &6.781                  &6.762                &6.743                  &6.747              &6.773          \\ \hline

States ($\mathcal{N} ^{2S+1}L_{J}$) &Present work      &\cite{comp-bc-2}       &\cite{comp-bc-6}          &\cite{comp-bc-7}   &\cite{comp-bc-8}      &\cite{comp-bc-9}      &\cite{comp-bc-10}  &\cite{Baldicchi:2000-bc-11}   &\cite{comp-bc-12}   \\\hline
$B_{c}$ (1 $^{1}S_{0}$)             &6.264          &6.271                  &6.286                      &6.310                   &6.255                 &6.280              &6.255               &6.258                 &6.28                          \\
$B_c^*$ (1 $^{3}S_{1}$)             &6.356          &6.338                  &6.341                      &6.355                   &6.320                 &6.321              &6.333               &6.334                 &6.35                             \\
$B_{c2}^*$  (1 $^3P_2$)             &6.814          &6.768                  &6.772                      &6.773                   &6.770                 &6.783                 &                   &                    &
\end{tabular}

\end{ruledtabular}
\footnotetext[1]{The CDF Collaboration confirms their earlier report
\cite{com-bc-exp-2} with higher statistical samples with a significance greater than 8$\sigma$ \cite{Bc-new-CDF}.}
% {The uncertainty weighted average of the results of Ref. \cite{com-bc-exp-1}, 6.40$\pm0.39\pm0.13$ GeV, and the results of Ref. \cite{com-bc-exp-2}, 6.286$\pm0.0053\pm0.0012$ GeV.}
% \bibitem{com-bc-exp-1}  F. Abe \emph{et al.} (CDF Collaboration),  Phys. Rev. Lett. \textbf{81}, 2432 (1998).
% \bibitem{com-bc-exp-2}  A. Abulencia \emph{et al.} (CDF Collaboration), Phys. Rev. Lett. \textbf{96}, 082002 (2006).
\end{table}

\subsubsection{Masses of the pure $s\bar{s}$ states}
   The masses of the pure $s\bar{s}$ states cannot be directly measured experimentally
because of the usual mixing of the pure isoscalar $n\bar{n}$ and
$s\bar{s}$ states.
  However, the comparison of the mass
of the pure $s\bar{s}$ state with that of the physical state can
help us to understand the mixing of the two isoscalar states of a
meson nonet.

  The masses of the pure $s\bar{s}$ states can be calculated from Eq. (\ref{high-equal-m}).
  When $i=n$, $j=s$, $k=b$ or $c$, inserting the corresponding masses into Eq. (\ref{high-equal-m}),
the masses of $s\bar{s}$ for the $1 ^{1}S_{0}$, $1 ^{3}S_{1}$ and
$1 ^{3}P_{2}$ multiplets are extracted and shown in Table 6.

  In Ref. \cite{Burakovsky-meson-relation}, under the approximation that mesons in the light quark sector have the common Regge slopes,
 two 6th power meson mass relations were derived to predict the masses of $c\bar{c}$ and $b\bar{b}$ meson states, respectively.
   Those two 6th power meson mass relations can be written as follows,
\begin{equation}
\begin{split}
&(M_{s\bar{s}}^2M_{n\bar{Q}}^2-M_{n\bar{n}}^2M_{s\bar{Q}}^2)(M_{s\bar{s}}^2-M_{n\bar{n}}^2)+M_{QQ}^2(M_{s\bar{Q}}^2-M_{n\bar{Q}}^2)(M_{s\bar{s}}^2-M_{n\bar{n}}^2) \\
=&
4(M_{s\bar{s}}^2M_{n\bar{Q}}^2-M_{n\bar{n}}^2M_{s\bar{Q}}^2)(M_{s\bar{Q}}^2-M_{n\bar{Q}}^2)
\label{6th-power},
\end{split}
\end{equation}
where $Q$ denotes $c$ or $b$.
   The results of applying Eq. (\ref{6th-power}) for the $1 ^{1}S_{0}$, $1 ^{3}S_{1}$ and $1 ^{3}P_{2}$ multiplets
to extract the masses of  the $s\bar{s}$ states are also shown in Table 6.
\begin{table}
    Table 6.   The masses of the pure $s\bar{s}$ states in pseudoscalar, vector and tensor meson multiplets
 given by Eqs. (\ref{high-equal-m}) and (\ref{6th-power}) (in units of GeV).\\
\begin{ruledtabular}

\begin{tabular}{c *{4}{c}}
$\mathcal{N} ^{2S+1}L_{J}$ &Eq. (\ref{high-equal-m}) i,j,k=n,s,c
&Eq. (\ref{high-equal-m}) i,j,k=n,s,b          &Eq.
(\ref{6th-power}) Q=c  &Eq. (\ref{6th-power}) Q=b     \\\hline

1 $^{1}S_{0}$                            &0.697  &0.698                         &0.761 or 0.157     &0.927 or 0.147   \\
1 $^{3}S_{1}$                            &1.009  &1.006                         &0.891 or 1.079     &0.841 or 1.145  \\

1 $^{3}P_{2}$                            &1.546  &1.544                         &1.492 or 1.582     &1.423 or 1.627   \\

\end{tabular}

\end{ruledtabular}
\end{table}

  From Table 5, one can see that the masses of $B_c$, $B_c^*$ and $B_{c2}^*$ given by Eq. (\ref{high-equal-factor})
 are bigger than those given in Refs. \cite{De-minLi2004,comp-bc-1,comp-bc-2,comp-bc-3,comp-bc-4,comp-bc-5,comp-bc-6,comp-bc-7,comp-bc-8,comp-bc-9,comp-bc-10,Baldicchi:2000-bc-11,comp-bc-12}.
   The mass of the $B_c$ meson given by Eq. (\ref{high-equal-m}) (present work)
 is better than those given by Eqs. (\ref{high-equal-factor}) and (\ref{14th-power}) comparing with experimental data.
   The masses of $B_c$, $B_c^*$ and $B_{c2}^*$ given by Eq. (\ref{high-equal-m}) (present work) are in reasonable
 agreement with those given in Refs. \cite{De-minLi2004,comp-bc-1,comp-bc-2,comp-bc-3,comp-bc-4,comp-bc-5,comp-bc-6,comp-bc-7,comp-bc-8,comp-bc-9,comp-bc-10,Baldicchi:2000-bc-11,comp-bc-12}.
% although the mass of $B_{c2}^*$ given by Eq. (\ref{high-equal-m}) is a little bigger ($\sim 0.05 GeV$).
   From Table 6, one can see that the masses of the pure $s\bar{s}$ state in the same multiplet
given by Eq. (\ref{high-equal-m}) are approximately the same when
we choose $k=c$ and $k=b$
 and they all satisfy the mass ranges shown in Table 3-1 which are given by the linear mass inequality (\ref{inequa1m}) and quadratic mass inequality (\ref{inequa2m}).
   However, the masses of the pure $s\bar{s}$ states given by Eq. (\ref{6th-power}) do not satisfy these constrains. % so good as those given by Eq. (\ref{high-equal-m}).

   As mentioned above, Eq. (\ref{6th-power}) was derived under the approximation that mesons in the light quark sector have the common Regge slopes
 and was applied for predicting the masses of charmonium and bottomonium \cite{Burakovsky-meson-relation}. % for the well-established multiplets with high accuracy.
   Obviously, Eq. (\ref{6th-power}) may be limited by this approximation while predicting the masses of light hadrons.
   Equation (\ref{14th-power}) was extracted under the same arguments on which Eq. (\ref{6th-power}) is based \cite{Burakovsky-meson-relation}.
   When $i=n$, $j=s$, and $k=Q$, Eq. (\ref{high-equal-m}) can be reduced to Eq. (\ref{6th-power}) if we choose $\frac{\alpha_{s\bar{s}}}{\alpha_{n\bar{n}}}=1$.
   Furthermore, with Eq. (\ref{high-equal-m}) one needs less meson states than those in the case of Eq. (\ref{14th-power}) to predict the masses of $\bar{b}c$ states.
   Therefore, Eq. (\ref{high-equal-m}) can properly describe the present meson spectroscopy \cite{PDG2006}.
%  Therefore, Eq. (\ref{high-equal-m}) is better than Eqs. (\ref{high-equal-factor}), (\ref{14th-power}) and (\ref{6th-power})
% for describing the present experimental data \cite{PDG2006}.

\subsection{Doubly charmed baryon $\Xi_{cc}^+(3520)$}

    The doubly charmed baryon $\Xi_{cc}^+(3520)$ (ccd) was first reported in the charged decay mode $\Xi_{cc}^+ \rightarrow \Lambda_c^+ K^- \pi^+$ (SELEX 2002)
 and confirmed in the decay mode $\Xi_{cc}^+ \rightarrow p D^+ K^-$ (SELEX 2005) \cite{Xi_cc}.
   These reports were adopted by the \emph{Particle Data Group} \cite{PDG2006} with the average mass 3518.9$\pm$0.9 $MeV$.
   However, the $J^P$ number has not been determined experimentally.
   Moreover, it has not been confirmed by other experiments (notably by BABAR \cite{Xi_cc(3520)-BABAR}, BELLE \cite{Xi_cc(3520)-BELLE} and FOCUS \cite{Xi_cc(3520)-FOCUS}),
% No evidence of this state is found by  BABAR \cite{Xi_cc(3520)-BABAR}, Belle \cite{Xi_cc(3520)-BELLE}, or FOCUS \cite{Xi_cc(3520)-FOCUS},
 even though they have O(10) (FOCUS) and O(100) (BABAR, BELLE) more reconstructed charm baryons than SELEX.
   This experimental puzzle  raised many theoretical discussions \cite{Xi_cc(3520)-not-3/2,Cohen-Liu-Martynenko-Edwards-Chang,Martynenko:2007je}.
   It was suggested that $\Xi_{cc}^+(3520)$ should be the ground state ($L=0$)
 with $J^P=\frac{1}{2}^+$ or $\frac{3}{2}^+$ due to its mass \cite{Xi_cc(3520)-not-3/2,Cohen-Liu-Martynenko-Edwards-Chang,Martynenko:2007je}.

  Now we will see whether the state $\Xi_{cc}^+(3520)$ could be assigned as a $\frac{3}{2}^+$ doubly charmed baryon.
Let us first assume that $\Xi_{cc}^+(3520)$ belongs to the $\frac{3}{2}^+$ multiplet.
  When $j=c$, $i=n$, and $q=n$, from Eq. (\ref{solution-b}), we have
\begin{equation}
\frac{\alpha_{\Xi_{cc}^*}'}{\alpha_{\Delta}'}= \frac{1}{2M_{\Xi_{cc}(3520)}^2}\times[(4M_{\Sigma_c^*}^2-M_{\Delta}^2-M_{\Xi_{cc}(3520)}^2)+\sqrt{(4M_{\Sigma_c^*}^2-M_{\Delta}^2-M_{\Xi_{cc}(3520)}^2)^2-4M_{\Delta}^2M_{\Xi_{cc}(3520)}^2} ],   \label{alpha_Xi-cc}
\end{equation}
\begin{equation}
\frac{\alpha_{\Sigma_c^*}'}{\alpha_{\Delta}'}= \frac{1}{4M_{\Sigma_c^*}^2}\times[(4M_{\Sigma_c^*}^2+M_{\Delta}^2-M_{\Xi_{cc}(3520)}^2)+\sqrt{(4M_{\Sigma_c^*}^2-M_{\Delta}^2-M_{\Xi_{cc}(3520)}^2)^2-4M_{\Delta}^2M_{\Xi_{cc}(3520)}^2} ]. \label{alpha_c}
\end{equation}
  When $j=c$, $i=s$, and $q=n$, from Eq. (\ref{solution-b}), we have
\begin{equation}
\frac{\alpha_{\Xi_{cc}^*}'}{\alpha_{\Xi^*}'}=  \frac{1}{2M_{\Xi_{cc}(3520)}^2}\times[(4M_{\Xi_c^*}^2-M_{\Xi^*}^2-M_{\Xi_{cc}(3520)}^2)+\sqrt{(4M_{\Xi_c^*}^2-M_{\Xi^*}^2-M_{\Xi_{cc}(3520)}^2)^2-4M_{\Xi^*}^2M_{\Xi_{cc}(3520)}^2} ] .  \label{alpha_Xi}
\end{equation}
  From Eq. (\ref{slope-whole}), we have
\begin{equation}
\frac{1}{\alpha_{\Delta}'}+\frac{2}{\alpha_{\Omega}'}=\frac{3}{\alpha_{\Xi^*}'}.  \label{alpha_omega}
\end{equation}
  Inserting the masses of $\Delta$,  $\Sigma_c^*$ and $\Xi_{cc}^+(3520)$ into Eq. (\ref{alpha_c}), we have
\begin{equation*}
\alpha_{\Sigma_c^*}'=0.867\alpha_{\Delta}'.
\end{equation*}
 Inserting the masses of $\Delta$,  $\Sigma_c^*$, $\Xi^*$,  $\Xi_c^*$ and $\Xi_{cc}^+(3520)$ into Eqs. (\ref{alpha_Xi-cc}) and (\ref{alpha_Xi}),
with the aid of Eq. (\ref{alpha_omega}), we have
\begin{equation*}
\alpha_{\Omega}'=0.860\alpha_{\Delta}'.
\end{equation*}

 Therefore, $\alpha_{\Omega}'\lesssim \alpha_{\Sigma_c^*}'$.
  This does not agree with the usual belief that the slopes of charmed baryons
 should be much smaller than the slopes of light noncharmed baryons.
  We have calculated the numerical results of $\frac{\alpha_{\Omega}'}{\alpha_{\Sigma_c^*}'}$
and find that it increases with the mass increase of $\Xi_{cc}^*$.
  Therefore, the mass of $\Xi_{cc}^*$ should be much bigger than the mass of $\Xi_{cc}^+(3520)$.
  In other words, the mass of $\Xi_{cc}^+(3520)$ is too small to be assigned as the $\frac{3}{2}^+$ doubly charmed baryons.

 According to the quark model, the lowest lying baryon states
should be the ground states ($L=0$) including the $J=\frac{1}{2}^+$ and $J=\frac{3}{2}^+$ doublets.   % the S-wave states, namely,
In the above discussion, we have manifested that the mass of $\Xi_{cc}^+(3520)$ is too small to be assigned as the $\frac{3}{2}^+$ doubly charmed baryons in Regge phenomenology.
   Therefore, we can conclude that
%  $ if $\Xi_{cc}^+(3520)$ is an S-wave doubly charmed baryon state it should have its $J^P$ as $\frac{1}{2}^+$.
$\Xi_{cc}^+(3520)$ should be the ground state with its $J^P$ as $\frac{1}{2}^+$.
   This assignment coincides with the fact that $\Xi_{cc}^+(3520)$ is observed to decay only weakly \cite{Xi_cc}
 (if the $J^P$ of $\Xi_{cc}^+(3520)$ were $\frac{3}{2}^+$, it should decay electromagnetically \cite{Xi_cc(3520)-not-3/2}).

    Inserting the masses of $\Sigma$, $\Xi$, $\Sigma_c$, $\Omega_c$ and $\Xi_{cc}^+(3520)$ into Eq. (\ref{Omega-cc}),
 we can get the mass of $\Omega_{cc}$, $M_{\Omega_{cc}}$$=3650.4 \pm 6.3 GeV$,
 where the uncertainty comes from the errors of the input data.
    Comparison of the masses of $\Xi_{cc}$ and $\Omega_{cc}$ extracted in the present work
 and those given in other references is shown in Table 7.

\begin{table}%[H]
 Table 7.   The masses of doubly and triply charmed baryons (in units of MeV).
  The numbers in boldface are the experimental values taken as the input. \\
\begin{ruledtabular}

\begin{tabular}{c*{5}{l}}
                                    & $\Xi_{cc}$                    & $\Omega_{cc}$                         & $\Xi_{cc}^*$              & $\Omega_{cc}^*$           & $\Omega_{ccc}$        \\ \hline
Pre.                         &\textbf{3518.9$\pm$0.9}        &3650.4$\pm$6.3                         &3684.4$\pm$4.4             &3808.4$\pm$4.3             &4818.9$\pm$6.8         \\

\cite{Burakovsky1997}              &$3610 \pm 3$                    &$3804 \pm 8$                       &$3735 \pm 17$                      &$3850 \pm 25$                &4930$\pm$45                       \\

\cite{Ponce-1979}                        &3511                      &3664                             &3630                     &3764                &4747                       \\

\cite{Vijande-etal-2004}         &3524  &3524     &3548  &3548       &4632               \\

\cite{Martynenko:2007je}            &3510 &3719 &3548 &3746 &4803      \\

\cite{H.R.Petry}         &3642             &3732         &3723            &3765        &4473   \\
\cite{Roberts:2007ni}     &3676               &3815    &3753                &3876            &4965  \\

\cite{Bjorken-1985}              &3635  &3800     &3695 $\pm$60  &3840$\pm$60        &4925$\pm$90               \\
\cite{Flynn-etal-2003}          &3549$\pm13 \pm$19$\pm$92          &3663$\pm11 \pm$17$\pm$95          &3641$\pm18 \pm$8$\pm$95       &3734$\pm14 \pm$8$\pm$97       &                       \\

\cite{Roncaglia:1995az}  &3660$\pm$70       &3740$\pm$80     &3740$\pm$70       &3820$\pm$80         &  \\
\cite{Ebert:2002-05-2007}  &3620         &3778       &3727          &3872            &     \\
\cite{He:2004px}                     &3520                          &3619                           &3630                           &3721           &       \\

\cite{Kiselev-etal-2000-2002}       &3478                           &3594                             &3610                                   &3730                    &                       \\

\cite{Martin-Richard-1995}       &  &3737           &   &3797               &4787               \\

\cite{Kiselev:2000jb}               &3550$\pm$80                &3650$\pm$80                    & &       &       \\

\cite{Jia-2006}                  &&&&       & 4760$\pm$60               \\
\cite{Hasenfratz:1980ka}        &&&&       & 4790               \\

\end{tabular}

\end{ruledtabular}
\end{table}

%2) Assuming that $\delta_{ij}^{1^{--}}$=$\delta_{ij}^{\frac{3}{2}^{+}}$, because of  \\   %gives $\Xi_{cc}=$3.650 GeV,           %But slopes $\alpha$ of heavy baryons not reasonable. Need further study.

\subsection{Parameters of Regge trajectories for the $\frac{3}{2}^{+}$ $SU(4)$ multiplet}

      In Ref. \cite{De-minLi2004}, the parameters of Regge trajectories for different meson multiplets and the masses of the meson states lying on those Regge trajectories were estimated.
  In this section, we will first extract the masses of the $\frac{3}{2}^{+}$ $SU(4)$ baryons absent from the baryon summary table so far.
  And then, with all the $\frac{3}{2}^{+}$ $SU(4)$ baryon masses and the value of $\alpha_\Delta$, we will calculate all the parameters (Regge slopes and intercepts) for the $\frac{3}{2}^{+}$ baryon trajectories.
  After that, we will estimate the masses of the orbital excited baryons lying on these Regge trajectories.

  All the masses of $\frac{3}{2}^{+}$ light baryons and charmed baryons are known experimentally.
  We need to know one of the masses of the baryons $\Xi_{cc}^*$, $\Omega_{cc}^*$ and $\Omega_{ccc}$
to calculate the masses of the other two states using the quadratic mass equalities (\ref{eq:whole}).
   First, we apply Eq. (\ref{equa-double}) to extract the mass of $\Xi_{cc}^*$ or $\Omega_{cc}^*$.
   When $i=n$, $j=c$, and $q=s$, we could insert the masses of $\Delta$, $\Sigma^*$, $\Xi^*$, $\Sigma_c^*$ and $\Xi_c^*$ into the relation (\ref{equa-double}) to calculate $M_{\Xi_{cc}^\ast}$.
   When $i=s$, $j=c$, and $q=s$, we could insert the masses of $\Sigma^*$, $\Xi^*$, $\Omega$, $\Xi_c^*$ and $\Omega_c^*$ into the relation (\ref{equa-double}) to calculate $M_{\Omega_{cc}^\ast}$.
%\begin{equation} \begin{split}
%& \frac{(4M_{\Xi^*}^2-M_{\Sigma^*}^2-M_{\Omega}^2)+\sqrt{(4M_{\Xi^*}^2-M_{\Sigma^*}^2-M_{\Omega}^2)^2-4M_{\Sigma^*}^2M_{\Omega}^2}}{2M_{\Omega}^2} \\
%=& \frac{[(4M_{\Xi_c^*}^2-M_{\Sigma^*}^2-M_{\Omega_{cc}^*}^2)+\sqrt{(4M_{\Xi_c^*}^2-M_{\Sigma^*}^2-M_{\Omega_{cc}^*}^2)^2-4M_{\Sigma^*}^2M_{\Omega_{cc}^*}^2}]/{2M_{\Omega_{cc}^*}^2}}{[(4M_{\Omega_c^*}^2-M_{\Omega}^2-M_{\Omega_{cc}^*}^2)+\sqrt{(4M_{\Omega_c^*}^2-M_{\Omega}^2-M_{\Omega_{cc}^*}^2)^2-4M_{\Omega}^2M_{\Omega_{cc}^*}^2}]/{2M_{\Omega_{cc}^*}^2}}  .
%\end{split} \end{equation}
    However, we find that the numerical results of $M_{\Xi_{cc}^\ast}$ and $M_{\Omega_{cc}^\ast}$ are very sensitive to the errors of the light baryon masses.
    Therefore, another way is needed to calculate the mass of $\Xi_{cc}^*$ or $\Omega_{cc}^*$.
   In Sec. III C, $\Xi_{cc}^+(3520)$ was assigned as the ground $\frac{1}{2}^{+}$ doubly charmed baryon.
   This may open a window to extract the masses of $\frac{3}{2}^{+}$ doubly charmed baryons.

   The first-order GMO formula for the baryon octet,
\begin{equation}
2(M_N+M_\Xi)=(3M_\Lambda+M_\Sigma)  \label{GMO-octet},
\end{equation}
is usually generalized to charmed cases by replacing $s$-quark with $c$-quark,
\begin{equation}
2(M_{N}+M_{\Xi_{cc}}) = 3M_{\Lambda_c}+M_{\Sigma_c} \label{n-Xi-cc-l}.
\end{equation}
 The quadratic form of Eq. (\ref{n-Xi-cc-l}) is
\begin{equation}
2(M_{N}^2+M_{\Xi_{cc}}^2) = 3M_{\Lambda_c}^2+M_{\Sigma_c}^2 \label{n-Xi-cc}.
\end{equation}
   However, the existence of high-order breaking effects in Eqs. (\ref{n-Xi-cc-l}) and (\ref{n-Xi-cc}) is obvious \cite{Burakovsky1997}.
   We use $\delta_{nc}^{\frac{1}{2}^+}$ to denote this effect in Eq. (\ref{n-Xi-cc}),
\begin{equation}
\delta_{nc}^{\frac{1}{2}^+}=M_{N}^2+M_{\Xi_{cc}}^2 - 2 (\frac{3M_{\Lambda_c}^2+M_{\Sigma_c}^2}{4}) .
\end{equation}

   Assuming that $\delta_{nc}^{\frac{1}{2}^+}$=$\delta_{nc}^{\frac{3}{2}^+}$, we have
\begin{equation}
\delta_{nc}^{\frac{1}{2}^+} =M_{N}^2+M_{\Xi_{cc}}^2 - 2 (\frac{3M_{\Lambda_c}^2+M_{\Sigma_c}^2}{4}) =\delta_{nc}^{\frac{3}{2}^+} =M_\Delta^2+M_{\Xi_{cc}^*}^2 - 2 M_{\Sigma_c^*}^2  \label{delta-nc-1232}.
\end{equation}
    Inserting the masses of $N$, $\Lambda_c$, $\Sigma_c$, $\Xi_{cc}^+(3520)$, $\Delta$ and $\Sigma_c^*$ into Eq. (\ref{delta-nc-1232}), we have
$M_{\Xi_{cc}^*}=3684.4 \pm4.4$ MeV,  where the uncertainty comes from the errors of the input data.

  Then, inserting the masses of $\Delta$, $\Omega$, $\Sigma_c^*$, $\Xi_c^*$ and $\Xi_{cc}^\ast$
into Eqs.  (\ref{equa-double}) and  (\ref{subeq:1}), we have
\begin{equation}
\begin{split}
& \frac{(4M_{\Sigma^*}^2-M_{\Delta}^2-M_{\Xi^*}^2)+\sqrt{(4M_{\Sigma^*}^2-M_{\Delta}^2-M_{\Xi^*}^2)^2-4M_{\Delta^*}^2M_{\Xi^*}^2}}{2M_{\Xi^*}^2} \\  \label{-Xicc}
=& \frac{[(4M_{\Xi_c^*}^2-M_{\Delta}^2-M_{\Xi_{cc}^*}^2)+\sqrt{(4M_{\Xi_c^*}^2-M_{\Delta}^2-M_{\Xi_{cc}^*}^2)^2-4M_{\Delta}^2M_{\Xi_{cc}^*}^2}]/{2M_{\Xi_{cc}^*}^2}}{[(4M_{\Xi_c^*}^2-M_{\Xi^*}^2-M_{\Xi_{cc}^*}^2)+\sqrt{(4M_{\Xi_c^*}^2-M_{\Xi^*}^2-M_{\Xi_{cc}^*}^2)^2-4M_{\Xi^*}^2M_{\Xi_{cc}^*}^2}]/{2M_{\Xi_{cc}^*}^2}}  ,
\end{split}
\end{equation}
\begin{equation}
M_{\Delta}^2+M_{\Xi^\ast}^2-2M_{\Sigma^\ast}^2=M_{\Sigma^\ast}^2+M_{\Omega}^2-2M_{\Xi^\ast}^2 .
\end{equation}
Then, we have the masses of $\Sigma^*$ and $\Xi^*$.
  Inserting the masses of $\Sigma^*$, $\Xi^*$, $\Omega$, $\Sigma_c^*$, $\Xi_c^*$ and $\Xi_{cc}^\ast$ into the quadratic mass equations in Eq. (\ref{eq:whole}),
we have the masses of $\Omega_c^*$, $\Omega_{cc}^\ast$ and $\Omega_{ccc}$.

    In this way, all the masses of $\frac{3}{2}^+$ $SU(4)$ baryons are known.
    With these masses and the value $\alpha_\Delta'=2/(M_{\Delta(1950)}^{2}-M_{\Delta}^{2})=0.9022$ $\pm$0.0285 GeV$^{-2}$
 (where the uncertainty comes from the errors of the input masses of $\Delta(1950)$ and $\Delta$),
  we have all the Regge slopes of $\frac{3}{2}^+$ trajectories from Eq. (\ref{solution-b}).
     Then, with these masses and the obtained Regge slopes,
 we have all the Regge intercepts of $\frac{3}{2}^+$ trajectories from Eq. (\ref{regge1}).

   From Eq. (\ref{regge1}), one has
\begin{equation}
M_{J+2}=\sqrt{M_J^2+\frac{2}{\alpha'}}  . \label{M-J+2}
\end{equation}
   Then, using this equation, the masses of the orbital excited baryons ($J^P=\frac{7}{2}^+,\frac{11}{2}^+$)
lying on the $\frac{3}{2}^+$ trajectories can be calculated.
  The Regge intercepts and the Regge slopes of the $\frac{3}{2}^+$ trajectories are shown in Table 8.
  The masses of light baryons,  charmed baryons, and doubly and triply charmed baryons lying on the $\frac{3}{2}^+$ trajectories are shown in Tables 9-1, 9-2 and 9-3, respectively.

\begin{table}
Table 8.  The Regge slopes (in units of $GeV^{-2}$) and the Regge intercepts of the $\frac{3}{2}^+$ trajectories. \\
\begin{ruledtabular}
{\small
\begin{tabular}{c *{10}{c}}
                    &$\Delta$       &$\Sigma^*$             &$\Xi^*$    &$\Omega$               &$\Sigma_c^*$       &$\Xi_{c}^*$    &$\Omega_{c}^*$                 &$\Xi_{cc}^*$    &$\Omega_{cc}^*$    &$\Omega_{ccc}$  \\\hline
$\alpha^\prime$     &0.902          &0.862              &0.825         &0.791                               &0.644            &0.623         &0.604                       &0.501         &0.488         &0.410  \\
                    &$\pm$0.029     &$\pm$0.036         &$\pm$0.042         &$\pm$0.047               &$\pm$0.023            &$\pm$0.026         &$\pm$0.029          &$\pm$0.019         &$\pm$0.021         &$\pm$0.016  \\\hline

\emph{a}(0)         &0.131    &-0.151        &-0.432         &-0.713                           &-2.583          &-2.864        &-3.145                                &-5.296        &-5.577        &-8.009    \\
                    &$\pm$0.046    &$\pm$0.074        &$\pm$0.102         &$\pm$0.133            &$\pm$0.147          &$\pm$0.174       &$\pm$0.203                   &$\pm$0.249        &$\pm$0.276        &$\pm$0.351    \\
\end{tabular}
}
\end{ruledtabular}
\end{table}
%\vspace{0.1cm}

\begin{table}%[H]
Table 9-1. The masses of the light baryons lying on the $\frac{3}{2}^+$ trajectories (in units of MeV).
The numbers in boldface are the experimental values taken as the input. \\
\begin{ruledtabular}

\begin{tabular}{r *{3}{l} *{3}{l}}
            &\multicolumn{3}{c}{$M_{\Delta}$}                                                                 &\multicolumn{3}{c}{$M_{\Sigma^*}$}  \\ \cline{2-4} \cline{5-7}
            &J=3/2&J=7/2&J=11/2                                                 &J=3/2&J=7/2&J=11/2 \\  \hline
Pre.    &\textbf{1232$\pm$1}   &\textbf{1932.5$\pm$17.5}  &2440$\pm$28                  &1383.9$\pm$2.3   &2058$\pm$22  &2560$\pm$36                   \\
Exp.    &1232$\pm$1       &1915$\sim$1950    &2300$\sim$2500                      &1384.6$\pm$2.6    &2015$\sim$2040   &                                                  \\
    %Exp.\cite{PDG2006}

\cite{H.R.Petry}        &1261   &1951   &2442         &1411   &2027   &     \\

\cite{Gonzalez:2006he}  &1232   &1921 &2175           &   &   &   \\
\cite{Klempt:2002tt}     &1232   &1950   &2467         &1394 &2056&       \\
\cite{Zhao:2006vk}      &1290    &1954   &               &1377  &2029   &            \\
\cite{Arndt:2003if}     &1232.9$\pm$1.2 &1923.3$\pm$0.5 &                                            \\

\cite{Capstick:1986-2000} &1230 &1940   &2450            &1370  &2060   &         \\
\cite{Isgur:1978wd}     &1240   &1915   &               &1390   &2015   &   \\  \hline

                                &\multicolumn{3}{c}{$M_{\Xi^*}$}                                                      &\multicolumn{3}{c}{$M_{\Omega}$}  \\        \cline{2-4} \cline{5-7}
                        &J=3/2&J=7/2&J=11/2                                                 &J=3/2&J=7/2&J=11/2\\  \hline
Pre.    &1530.2$\pm$1.9    &2183$\pm$27   &2681$\pm$45                          &\textbf{1672.45}$\pm$\textbf{0.29}        &2308$\pm$32   &2802$\pm$54 \\
Exp.    &1533.4$\pm$2.1      & &                                                       &1672.45$\pm$0.29 & &  \\
    % Exp.\cite{PDG2006}

\cite{H.R.Petry}        &1539   &2169   &         &1636   &2292   &     \\
\cite{Gonzalez:2006he}  &&&           &   &   &   \\
\cite{Klempt:2002tt}    &1540 &2157 &           &1672   &   &           \\
\cite{Zhao:2006vk}      &1502   &2142   &               &1665   &2293&              \\
\cite{Arndt:2003if}                                                 \\

\cite{Capstick:1986-2000} &1505 &2180   &            &1635  &2295   &         \\
\cite{Isgur:1978wd}     &1530   &&              &1675    &&\\

\end{tabular}

\end{ruledtabular}
\end{table}
%\vspace{0.1cm}

\begin{table}%[H]
Table 9-2. The masses of the charmed baryons lying on the $\frac{3}{2}^+$ trajectories (in units of MeV).
The numbers in boldface are the experimental values taken as the input.\\
\begin{ruledtabular}

\begin{tabular}{r *{3}{l} *{3}{l} *{3}{l}}

                            &\multicolumn{3}{c}{$M_{\Sigma_c^*}$}                                               &\multicolumn{3}{c}{$M_{\Xi_c^*}$}                                                                      &\multicolumn{3}{c}{$M_{\Omega_c^*}$}\\    \cline{2-4} \cline{5-7} \cline{8-10}
        &J=3/2&J=7/2&J=11/2                                             &J=3/2&J=7/2&J=11/2                                                           &J=3/2&J=7/2&J=11/2\\  \hline
Pre.    &\textbf{2518.0$\pm$1.9}  &3073$\pm$18  &3543$\pm$30         &\textbf{2646.4$\pm$1.6}   &3196$\pm$22    &3664$\pm$37             &2774.1$\pm$5.5   &3318$\pm$28   &3784$\pm$46\\
Exp.    &2518.0$\pm$1.9     & &                                         &2646.6$\pm$1.4 & &                                                        &2768.3$\pm$3 & & \\
% Exp.\cite{PDG2006}

\cite{Ponce-1979} &2481    &    &                 &2642    &    &                 & 2764   &    &       \\

\cite{H.R.Petry}        &2539   & &           &2651   & &       &2721   & &\\
\cite{Roberts:2007ni}   &2519   &3015   &               &2650   &3100   &           &2776   &3206   &\\

\cite{Roncaglia:1995az} &2520$\pm$20   &   &              &2650$\pm$20   & &               &2770$\pm$30 & &  \\
\cite{Ebert:2002-05-2007} &2518 &3015   &               &2654   &3136   &           &2768   &3237   &   \\

\cite{Capstick:1986-2000} &2495 &3090   &            & &  &         \\
\cite{Isgur:1978wd}     &2510   &3010   &              &    &&\\

\end{tabular}

\end{ruledtabular}
\end{table}
%\vspace{0.1cm}

\begin{table}
Table 9-3. The masses of the doubly and triply charmed baryons lying on the $\frac{3}{2}^+$ trajectories (in units of MeV).\\
\begin{ruledtabular}

\begin{tabular}{r *{3}{l} *{3}{l} *{3}{l}}
                    &\multicolumn{3}{c}{$M_{\Xi_{cc}^*}$}                   &\multicolumn{3}{c}{$M_{\Omega_{cc}^*}$}                            &\multicolumn{3}{c}{$M_{\Omega_{ccc}}$}\\   \cline{2-4} \cline{5-7}  \cline{8-10}
                &J=3/2&J=7/2&J=11/2                         &J=3/2&J=7/2&J=11/2                                 &J=3/2&J=7/2&J=11/2\\ \hline
Pre.    &3684.4$\pm$4.4 &4192$\pm$19 &4644$\pm$32             &3808.4$\pm$4.3 &4313$\pm$23 &4765$\pm$39             &4818.9$\pm$6.8 &5302$\pm$21 &5744$\pm$34  \\
Exp.    &&&             &&&             &&& \\
% Exp. \cite{PDG2006}

\cite{Roberts:2007ni}   &3753   &4097   &               &3876   &4230   &           &4965   &5331   &\\
\cite{Kiselev-etal-2000-2002}     &3610  &4089  &             &3730 &&                         & & & \\

%Ref.\cite{Ponce-1979} &3630    &    &             &3764    &    &             &4747    &    &       \\
%Ref.\cite{H.R.Petry}        &3723   &   &         &3765   &   &            &4473   &   &\\
%Ref.\cite{Ebert:2002-05-2007} &3727 &   &           &3872   &   &           &   &   &   \\

%Ref.\cite{Roncaglia:1995az} &3740$\pm$70   &   &         &3820$\pm$80   &&                       &&&              \\

\end{tabular}

\end{ruledtabular}
\end{table}

   The masses of $\Xi_{cc}^*$, $\Omega_{cc}^\ast$ and $\Omega_{ccc}$ extracted in the present work
and those given in other references are also shown in Table 7.
   From Table 7, we can see that the masses of $\frac{1}{2}^+$ and $\frac{3}{2}^+$ doubly and triply charmed baryons predicted by us
agree well with those given in most other references. % except Ref. \cite{Burakovsky1997}
   The predictions in Ref. \cite{Burakovsky1997} are bigger than ours because of the approximation adopted there that
baryons in the light quark sector have common Regge slopes.
  The mass splitting obtained in the framework of nonrelativistic effective field theories of QCD, $M_{\Xi_{cc}^*}-M_{\Xi_{cc}}=120 \pm40 MeV$ (see Ref. \cite{Brambilla-Vairo-2005} and references therein),
agrees with our present results shown Table 7.

%   Thereafter, if we know the mass for one of the doubly or triply bottom baryons,
% with the masses of light baryons, charmed baryons, and $\Sigma_b^{\ast}$, using the quadratic mass equalities (\ref{eq:whole}),
% we can have all the masses of baryons containing b quark and belonging to the $\frac{3}{2}^+$ multiplet.
%Now, how to determine the states $\frac{3}{2}^{+}$ $\Xi_{cc}$ (or $\Omega_{cc}$, $\Omega_{ccc}$) and $\Xi_{bb}$ (or $\Omega_{bb}$, $\Omega_{bbb}$)?

%Let $\Xi_{bb}=$  [average of Lattice QCD]

\subsection{Parameters of Regge trajectories for the $\frac{1}{2}^{+}$ $SU(4)$ multiplet}

    Up to now, all the masses of ground $\frac{1}{2}^+$ $SU(4)$ baryons are known.
    We will determine the Regge slopes and intercepts of the $\frac{1}{2}^{+}$ $SU(4)$ multiplet
and give predictions for masses of the $\frac{5}{2}^{+}$ and $\frac{9}{2}^{+}$ baryon states lying on these Regge trajectories.

   Recently, the spin-parity of the $\Lambda_c^+(2880)$ baryon was determined by experiment.
     $\Lambda_c^+(2880)$ was observed by CLEO in the $\Lambda_c\pi^+\pi^-$ mode \cite{Lambda_c2880-CLEO}
 and then confirmed by BABAR in the $D^0p$ mode recently \cite{Lambda_c2880-2940b}.
  From the analysis of the angular distribution in its $\Sigma_c(2455)\pi$ decays
and the small ratio, $\Gamma_{\Sigma_c(2520)\pi} / \Gamma_{\Sigma_c(2455)\pi}$ $\backsimeq$ $0.23$, measured by BELLE
it is concluded that the $J^P$ of $\Lambda_c^+(2880)$ is $\frac{5}{2}^+$ \cite{Lambda_c2880-2940a}.
  This spin-parity assignment is in agreement with the theoretical investigation
that $\Lambda_c^+(2880)$ is the orbital ($L=2$) excitation of $\Lambda_c^+$ \cite{Ebert:2002-05-2007,Cheng-Chua}. % (analogous to $\Lambda(1820)$)
 Therefore, $\Lambda_c^+(2880)$ and $\Lambda_c^+$ lie on the common Regge Trajectory.
 We can have the Regge slope of $\Lambda_c^+$ from Eq. (\ref{alpha}),
\begin{equation}
\alpha_{\Lambda_c}'=\frac{\frac{5}{2}-\frac{1}{2}}{M_{\Lambda_c^+(2880)}^2-M_{\Lambda_c^+}^2}=0.650 \pm0.005 \hspace{0.1cm} GeV^{-2} .
\end{equation}
   From Eq. (\ref{alpha}), we also have
\begin{equation*} \alpha_N'=\frac{2}{M_{N(1680)}^2-M_{N}^2}=1.022\pm0.009 \hspace{0.1cm} GeV^{-2}, \end{equation*}
\begin{equation} \alpha_\Lambda'=\frac{2}{M_{\Lambda(1820)}^2-M_{\Lambda}^2}=0.967\pm0.009 \hspace{0.1cm} GeV^{-2}.  \end{equation}
   We assume that       $\alpha_{\Sigma}'=\alpha_{\Sigma^*}'$,  $\alpha_{\Xi}'=\alpha_{\Xi^*}'$,
        $\alpha_{\Sigma_c}'=\alpha_{\Sigma_c^*}'$,   $\alpha_{\Xi_c'}'=\alpha_{\Xi_c^*}'$,   $\alpha_{\Omega_c}'=\alpha_{\Omega_c^*}'$,
        $\alpha_{\Xi_{cc}}'=\alpha_{\Xi_{cc}^*}'$,   and $\alpha_{\Omega_{cc}}'=\alpha_{\Omega_{cc}^*}'$.
%     As mentioned in Subsection C of Sec. II, the value of $\alpha_{\Xi_c}'$ is slightly bigger than that of $\alpha_{\Xi_c'}'$.
   Although the slopes of a heavy baryon containing a scalar diquark and that containing an axial-vector diquark are different,
 we assume that $\gamma_s$ for the heavy baryons containing scalar diquarks
 is approximately the same as $\gamma_s$ for heavy baryons containing axial-vector diquarks, \emph{i.e.},
 $\frac{1}{\alpha_{\Xi_c}'}-\frac{1}{\alpha_{\Lambda_c}}=\frac{1}{\alpha_{\Xi_c'}'}-\frac{1}{\alpha_{\Sigma_c}}$.
    Then, all the Regge slopes of $\frac{1}{2}^+$ $SU(4)$ baryons are known and shown in Table 10.

\begin{table}
Table 10.   The Regge intercepts and Regge slopes of the $\frac{1}{2}^+$ trajectories.
\begin{ruledtabular}
{\small
\begin{tabular}{c *{11}{c}}

            &$N$   &$\Lambda$&$\Sigma$   &$\Xi$              &$\Lambda_c$&$\Sigma_c$   &$\Xi_c$&$\Xi_c^\prime$   &$\Omega_c$    &$\Xi_{cc}$  &$\Omega_{cc}$ \\ \hline
\emph{a}(0) &-0.401 &-0.704&-0.727   &-0.933                         &-2.900 &-3.377   &-3.337&-3.638     &-3.892                        &-5.699 &-6.002\\
  & $\pm$0.010 &$\pm$0.011 &$\pm$0.059    & $\pm$0.082          &$\pm$0.003 &$\pm$0.137   &$\pm$0.043 &$\pm$0.184   &$\pm$0.217        &$\pm$0.228   &$\pm$0.291  \\\hline

$\alpha^\prime$ &1.022  &0.967 &0.862    &0.825              &0.650 &0.644   &0.629 &0.623    &0.604                                      &0.501  &0.488  \\
  &$\pm$0.009   &$\pm$0.009 &$\pm$0.036   &$\pm$0.042        &$\pm$0.005 &$\pm$0.022   &$\pm$0.006 &$\pm$0.026     &$\pm$0.029          &$\pm$0.018  &$\pm$0.020  \\
\end{tabular}
}
\end{ruledtabular}
\end{table}

    % After that,
    With the masses and the obtained Regge slopes for the $\frac{1}{2}^+$ baryons,
 we have all the Regge intercepts of $\frac{1}{2}^+$ trajectories from Eq. (\ref{regge1}).
   Then, using Eq. (\ref{M-J+2}), the masses of orbital excited baryons ($J^P=\frac{5}{2}^+$, $\frac{9}{2}^+$)
lying on the $\frac{1}{2}^+$ trajectories can be calculated.
  The Regge intercepts of the $\frac{1}{2}^+$ trajectories are also shown in Table 10.
  The masses of light baryons,  charmed baryons and doubly charmed baryons lying on the $\frac{1}{2}^+$ trajectories are shown in Tables 11-1, 11-2 and 11-3, respectively.

\begin{table}%[H]
Table 11-1. The masses of the light baryons lying on the $\frac{1}{2}^+$ trajectories (in units of MeV).
The numbers in boldface are the experimental values taken as the input. \\
\begin{ruledtabular}
{\tiny  % \scriptsize
\begin{tabular}{r lll lll lll lll}
    &\multicolumn{3}{c}{$M_N$}    &\multicolumn{3}{c}{$M_{\Lambda}$}   &\multicolumn{3}{c}{$M_{\Sigma}$}    &\multicolumn{3}{c}{$M_{\Xi}$}    \\   \cline{2-4}  \cline{5-7}  \cline{8-10}  \cline{11-13}
     &   J=1/2&J=5/2&J=9/2&     J=1/2&J=5/2&J=9/2&      J=1/2&J=5/2&J=9/2&    J=1/2&J=5/2&J=9/2\\\hline

Pre. &\textbf{938.92}&\textbf{1685}&2190                  &\textbf{1115.683}&\textbf{1820} &2319          &\textbf{1193.17}&1935&2463           &\textbf{1318.07}&2040&2566\\
      &$\pm$\textbf{0.65} &$\pm$\textbf{5} &$\pm$8.0    &$\pm$\textbf{0.006} &$\pm$\textbf{5} &$\pm$7.8    &$\pm$\textbf{4.11} &$\pm$27 &$\pm$41    &$\pm$\textbf{4.31} &$\pm$33 &$\pm$50  \\\hline

Exp. &938.92    &1680       & 2200                   &1115.683 &1815 &2340                           &1193.17 &1900 &                        &1318.07 &2025 & \\
      &$\pm$0.65 &$\sim$1690  &$\sim$2300             &$\pm$0.006 &$\sim$1825 &$\sim$2370                 &$\pm$4.11 &$\sim$1935 &                &$\pm$4.31    &$\pm$5  &  \\

\cite{H.R.Petry}        &939  &1723 &2221         &1108  &1834   &2340     &1190 &1956  &           &1310   &2013   &\\

\cite{Gonzalez:2006he}  &940 &1722  &2378           &   &   &   \\

\cite{Klempt:2002tt}     &939 &1779  &2334          &1144 &1895 &2424         &1144 &1895 &2424            &1317   &2004   &2510   \\

\cite{Zhao:2006vk}      &990 &1744  &         &1115 &1844 &         &1192 &1906 &        &1317 &2014 &            \\

\cite{Arndt:2003if}     &   &1683.2$\pm$0.7 &2270$\pm$11                                             \\

\cite{Capstick:1986-2000} &960 &1770   &2345            &1115  &1890  &        &1190   &1955  &           &1305   &2045   &  \\

\cite{Isgur:1978wd}         &940 &1715 &         &1110 &1815 &         &1915 &1940 &        &1320&&      \\

\end{tabular}
}
\end{ruledtabular}
\end{table}

\begin{table}%[H]
Table 11-2.  The masses of the charmed baryons lying on the $\frac{1}{2}^+$ trajectories (in units of MeV).
The numbers in boldface are the experimental values taken as the input. \\
\begin{ruledtabular}
{\tiny  % \scriptsize
\begin{tabular}{r lll lll lll lll lll}
    &\multicolumn{3}{c}{$M_{\Lambda_c}$}    &\multicolumn{3}{c}{$M_{\Sigma_c}$}                                  &\multicolumn{3}{c}{$M_{\Xi_c}$}       &\multicolumn{3}{c}{$M_{\Xi_c^\prime}$}                      &\multicolumn{3}{c}{$M_{\Omega_c}$}    \\  \cline{2-4}  \cline{5-7}  \cline{8-10}  \cline{11-13}  \cline{14-16}
  &J=1/2&J=5/2&J=9/2  &J=1/2&J=5/2&J=9/2                                                             &J=1/2&J=5/2&J=9/2  &J=1/2&J=5/2&J=9/2                                             &J=1/2&J=5/2&J=9/2\\\hline

Pre. &\textbf{2286.46} &\textbf{2881.5} &3737     &\textbf{2453.56} &3021 &3497                       &\textbf{2469.5} &3046 &3529         &\textbf{2576.9} &3138 &3614               &\textbf{2697.5} &3254 &3729 \\
  &$\pm$\textbf{0.14}&$\pm$\textbf{0.3} &$\pm$0.61    &$\pm$\textbf{0.85} &$\pm$18 &$\pm$31        &$\pm$\textbf{2.0} &$\pm$7 &$\pm$10    &$\pm$\textbf{4.2} &$\pm$24 &$\pm$40        &$\pm$\textbf{2.6} &$\pm$26  &$\pm$44     \\\hline

Exp. &2286.46 &2881.5 &                      &2453.56 & &                                &2469.5    &    &          &2576.9 &  &                   &2697.5     &   & \\
            &$\pm$0.14      &0.3 &                                      &$\pm$0.85 & &                      &$\pm$1.2  &    &          &$\pm$4.2 & &                              &$\pm$2.6   &   &     \\
% Exp.\cite{PDG2006}
\cite{Ponce-1979} &2243   &&                      &2380  &&                           &2425  &  &                           &2530  &  &                       &2678  & &   \\

\cite{H.R.Petry}        &2272   &   &                   &2459   &   &                           &2469   &   &                       &2595   &   &                   &2688   &   &       \\
\cite{Roberts:2007ni}   &2268   &2887   &               &2455   &3003   &                       &2492   &2995   &                   &2592   &3100 &             &2718   &3196 & \\

\cite{Roncaglia:1995az} &2285$\pm$1   &   &               &2453$\pm$3 &&                            &2468$\pm$3    &&                         &2580$\pm$20    &&           &2710$\pm$30&&   \\
\cite{Ebert:2002-05-2007} &2294 &2883   &               &2439   &2960   &                       &2481   &3042   &                   &2578   &3087 &             &2698   &3187 & \\

\cite{Capstick:1986-2000} &2265 &2910  &                &2440  &3065   &         \\
\cite{Isgur:1978wd}     &2260   &2810 &              &2440    &3010 &\\

\end{tabular}
}
\end{ruledtabular}
\end{table}

\begin{table}
Table 11-3. The masses of the doubly charmed baryons lying on the $\frac{1}{2}^+$ trajectories (in units of MeV).
The numbers in boldface are the experimental values taken as the input. \\
\begin{ruledtabular}

\begin{tabular}{r lll lll}
  &\multicolumn{3}{c}{$M_{\Xi_{cc}}$}                   &\multicolumn{3}{c}{$M_{\Omega_{cc}}$}                \\  \cline{2-4}  \cline{5-7}
  &J=1/2&J=5/2&J=9/2                                    &  J=1/2&J=5/2&J=9/2        \\\hline
Pre. &\textbf{3518.9$\pm$0.9}       &4047$\pm$19 &4514$\pm$33         &3650.4$\pm$6.3 &4174$\pm$26 &4639$\pm$41 \\
Exp.     &3518.9$\pm$0.9      &    &                         & & & \\   %\cite{PDG2006}

\cite{Roberts:2007ni}   &3676   &4047   &               &3815   &4202   &     \\
\cite{Kiselev-etal-2000-2002}     &3478  &4050  &                         &3594 & & \\

%Ref.\cite{Ponce-1979} &3511    &    &              &3664    &    &       \\
%Ref.\cite{H.R.Petry}        &3642   &   &               &3732   &   &      \\
%Ref.\cite{Ebert:2002-05-2007} &3620 &    &               &3778   &    &     \\
%Ref.\cite{Roncaglia:1995az} &3660$\pm$70   &   &         &3740$\pm$80 &&       \\
\end{tabular}

\end{ruledtabular}
\end{table}

\subsection{Charm-strange baryons }
  There are  five charm-strange baryons presented in PDG 2006 \cite{PDG2006}: $\Xi_c$, $\Xi_c'$, $\Xi_c^\ast$, $\Xi_c(2790)$ and $\Xi_c(2815)$.
$\Xi_c(2790)$ and $\Xi_c(2815)$ were assigned as the first orbital (1P) excitations of $\Xi_c$
with $J^P=\frac{1}{2}^-$ and $J^P=\frac{3}{2}^-$, respectively.

  Recently, $\Xi_c(2980)$ and $\Xi_c(3077)$ were
first reported by BELLE \cite{Xi_2980-Xi_3077b} and then confirmed by BABAR \cite{Xi_2980-Xi_3077a}.
 BABAR also reported the observation of $\Xi_c^+(3055)$ and $\Xi_c^+(3123)$ \cite{xi_3055-3123-babar}.
 The $J^P$ of $\Xi_c(2980)$, $\Xi_c (3055)$, $\Xi_c(3077)$ and $\Xi_c (3123)$ have not been measured.
 The masses of these states imply that they could be the states with the total quark orbital angular momentum $L=2$.
Here we attempt to study which Regge trajectory these states may lie on. % the $J^P$ assignment of these states in Regge trajectory.

  From Table 11-2, it can be seen that the mass of $\Xi_c(3123)$ coincides with the mass of $\Xi_c'(\frac{5}{2}^+)$.
Therefore, $\Xi_c(3123)$ probably lies on the Regge trajectory of $\Xi_c'$.
 In other words, $\Xi_c(3123)$ may be the orbital excited ($J^P=\frac{5}{2}^+$) state of $\Xi_c'$ containing an axial-vector diquark.
This assignment is in agreement with Ebert's assignment in the relativistic quark model \cite{Ebert:2002-05-2007}.
   We can also see that both the masses of  $\Xi_c(3055)$ and $\Xi_c(3077)$ are near the mass of $\Xi_c(\frac{5}{2}^+)$.
% 2) Both $\Xi_c(3055)$ and $\Xi_c(3077)$ may be the orbital excited ($L=2$) states of $\Xi_c$ containing scalar diquarks, namely, the $J^P=$ ($\frac{3}{2}^+$, $\frac{5}{2}^+$) partners.
%   The width of the $J^P=$ $\frac{3}{2}^+$ member of this partners is broader than the width of the $J^P=$ $\frac{5}{2}^+$ member according to the analysis on the decay models \cite{Rosner2007width}.
%   It was measured that $\Gamma_{\Xi_c(3055)} \simeq $ 17 MeV \cite{xi_3055-3123-babar}, $\Gamma_{\Xi_c(3077)} \simeq $ 6  MeV \cite{Xi_2980-Xi_3077a, Xi_2980-Xi_3077b}.
    The mass of $\Xi_c(2980)$ is lower compared with that of $\Xi_c(\frac{5}{2}^+)$ or $\Xi_c'(\frac{5}{2}^+)$.
% 3) $\Xi_c(2980)$ may not lie on the Regge trajectory of $\Xi_c$ and $\Xi_c'$.

% remarks comments recounts observations
  The above comments can be seen more clearly when combining with the slopes of these baryons.
   As mentioned above, the slopes of Regge trajectories decrease with quark mass increase.
Therefore, the slope of $\Xi_c$ ($\Xi_c'$, $\Xi_c^\ast$) is less than the slope of $\Lambda_c$,
\begin{equation}
\alpha_{\Xi_c^{(\prime,*)}}'< 0.650 \hspace{0.1cm} GeV^{-2}.  \label{xic-less-lambdac}
\end{equation}

  Assuming that $\Xi_c(2980)$, $\Xi_c (3055)$, $\Xi_c(3077)$ or $\Xi_c(3123)$
lies on the same Regge trajectory with $\Xi_c^{(\prime,*)}$, respectively,
so that the difference between the angular momenta of these baryons with those of $\Xi_c^{(\prime,*)}$ is $\Delta L= 2$,
 we obtain the values of the Regge slopes for $\Xi_c^{(\prime,*)}$ shown in Table 12.
\begin{table}
    Table 12.    The values (in units of GeV $^{-2}$) of the Regge slope for $\Xi_c^{(\prime,*)}$ given from Eq. (\ref{regge1}) under the assumption that
$\Xi_c(2790)$, $\Xi_c(2815)$, $\Xi_c^+(3055)$ or $\Xi_c(3123)$
lies on the same Regge trajectory with $\Xi_c^{(\prime,*)}$, respectively.   \\
\begin{ruledtabular}

\begin{tabular}{r *{4}{l}}
                    &$\Xi_c(2980)$  &$\Xi_c(3055)$  &$\Xi_c(3077)$  &$\Xi_c(3123)$  \\\hline
$\alpha_{\Xi_c}'$   &0.728         &0.619         &0.591         &0.547         \\
$\alpha_{\Xi_c'}'$  &0.907         &0.944         &0.703         &0.643       \\
$\alpha_{\Xi_c^\ast}'$ &1.086      &0.860         &0.806         &0.727       \\
\end{tabular}

\end{ruledtabular}
\end{table}

       From the relation (\ref{xic-less-lambdac}), Table 10, Table 11-2 and Table 12, we can conclude that:
   $\Xi_c(2980)$ cannot lie on the Regge trajectory of $\Xi_c$, $\Xi_c'$ and $\Xi_c^\ast$.
 ($\Xi_c(2980)$ can be interpreted in the relativistic quark model as the first radial (2S) excitation of the $\Xi_c$  with $J^P=\frac{1}{2}^+$ containing the light axial-vector diquark \cite{Ebert:2002-05-2007}.)
%
%   We know that $\alpha_{\Xi_c}' >\alpha_{\Xi_c^{\prime}}' >\alpha_{\Xi_c^{*}}'=0.623\pm0.026 \hspace{0.1cm} GeV^{-2}$.
   Both $\Xi_c(3055)$ and $\Xi_c(3077)$ can be assigned as the $J^P=$ $\frac{5}{2}^+$ state. % lying on the Regge trajectory of $\Xi_c$.   % ($\Xi_c(3077)$ cannot be excluded).
   $\Xi_c(3123)$ probably lies on the Regge trajectory of $\Xi_c'$.  In other words, $\Xi_c(3123)$ may be the orbital excited ($\Delta$L=2) state of $\Xi_c'$ with $J^P=\frac{5}{2}^+$ containing an axial-vector diquark.
   Further study is needed to determine the $J^P$ of these states more accurately.

%\newpage
\section{Discussion  and Conclusion}

%data check relations (\ref{3/2equation2})

  In this work, under the main assumption that the quasilinear Regge trajectory ansatz is suitable to
describe meson spectra and baryon spectra, with the requirements of the additivity of intercepts and inverse slopes,
 some useful linear mass inequalities, quadratic mass inequalities and quadratic mass equalities are
derived for mesons and baryons.

   Based on these relations, we have given upper limits and lower limits for some mesons and baryons.
  The masses of $\bar{b}c$ and $s\bar{s}$ belonging to the pseudoscalar ($1 ^1S_0$), vector ($1 ^3S_1$) and tensor ($1 ^3P_2$) meson multiplets are also extracted.
    We suggest that the $J^P$ of $\Xi_{cc}^+(3520)$ should be $\frac{1}{2}^+$.  % if it is an S-wave state.
    % We suggest that $\Xi_{cc}^+(3520)$ should be the ground state with $J^P=\frac{1}{2}^+$.
  The numerical values for the parameters of the $\frac{1}{2}^+$ and $\frac{3}{2}^+$ $SU(4)$ baryon trajectories are extracted and
the masses of the orbital excited baryons lying on the $\frac{1}{2}^+$ and $\frac{3}{2}^+$
trajectories are estimated.
   We propose that $\Xi_c(3123)$ may be a candidate for the orbital excited ($\Delta$L=2) state of $\Xi_c'$
with $J^P=\frac{5}{2}^+$ containing an axial-vector diquark.
%  $\Xi_c(3055)$ may be a candidate for the orbital excited ($\Delta$L=2) state of $\Xi_c$
% with $J^P=\frac{5}{2}^+$ containing a scalar diquark, and
 The predictions are in reasonable
agreement with the existing experimental data and those suggested in many other
different approaches.
\\

%  Besides, Moreover, In addition, Encouragingly,
   In Sec. II C, we showed that the linear mass GMO formula is an inequality in fact and the quadratic mass GMO formula is also an inequality with the sign opposite to the linear case.
  Encouragingly, the linear meson mass inequalities (\ref{inequa1m}) and the linear baryon mass inequalities (\ref{inequa1b})
 are similar to those derived from a general
 illation in QCD for the ground hadron states \cite{inequalities-QCD1,inequalities-QCD2,inequalities-QCD3}
 (The authors of Ref. \cite{inequalities-QCD3} also point out that
 the linear mass inequalities (\ref{inequa1m}) and (\ref{inequa1b}) hold for many potentials,
 although the linear baryon mass inequality (\ref{inequa1b}) does not hold for some special potentials).
%
%    The inequalities in Refs. \cite{inequalities-QCD1,inequalities-QCD2} include equal sign.
%    In Subsection C of Sec. II, we argued that the equal sign in is not necessary in the inequalities (\ref{inequa1m0}) for $i\neq j$ in Regge theory.
%
%
    In Ref. \cite{inequalities-QCD2}, Nussinov and Lampert showed that
 the linear meson mass inequality (\ref{inequa1m}) satisfies the experimental data of the well-established meson multiplets
 (vector $1 ^3S_1$, tensor $1 ^3P_2$, axial-vector $1 ^3P_1$ and scalar $1 ^3P_0$) with different flavor combinations of $i$ and $j$,
 and the linear baryon mass inequality (\ref{inequa1b}) satisfies the experimental data of the baryon octet and the baryon decuplet.
    They gave the lower limits for the masses of some unobserved mesons and baryons with the linear mass inequalities.
% (Upper limits of these unobserved hadrons were not given by them).
%
%
   In our work, in addition to the lower limits, we also give the upper limits for the masses of hadrons.
 We can see from Table 3-1, 3-2 and 4 that these limits agree with the existing data.
 The mass ranges in Table 3 and 4 are narrow (smaller than 0.5 GeV) for hadrons which do not contain $b$-quark.
 These mass ranges will be useful for the discovery of the unobserved hadron states. % and the spin-parity assignment of these states.
    When $b$-quark is involved, the mass ranges in Tables 3 and 4 become large (could be as large as 1 to 2 GeV)
 and consequently, the constraints become weaker.
   However, since many hadrons containing $b$-quark have not been observed in experiments,
 these mass ranges may also provide helpful guidance for the discovery of these hadrons.

   As far as we know, there is only one work to study the quadratic meson mass inequalities.
 %  There have been few works to study quadratic mass inequalities.
 % We only find one work to study the quadratic meson mass inequalities.
  In Ref. \cite{inequalities-history}, with the current-algebra technique,
 corrections to the GMO quadratic mass formula due to second-order SU(4) breaking was discussed by Simard and Suzuki.
   They gave a quadratic mass inequality for pseudoscalar mesons,
\begin{equation}
\frac{1}{2}\left[M_\pi^2+\left(\frac{2}{3}M_\eta^2+\frac{1}{3}M_{\eta'}^2\right)\right]+M_{\eta_c(1S)}^2-2M_{D}^2
> 0 , \label{Simart-r0}
\end{equation}
 and two quadratic mass inequalities for vector mesons,
\begin{equation}
\frac{1}{2}(M_\rho^2+M_\omega^2)+M_{J/\psi(1S)}^2-2M_{D^*}^2 < 0 ,
\label{Simart-r1}
\end{equation}
\begin{equation}
M_\phi^2+M_{J/\psi(1S)}^2-2M_{D_s^*}^2 < 0 . \label{Simart-r2}
\end{equation}

  The sign of the quadratic mass inequality (\ref{Simart-r0}) is the same as that of our quadratic mass inequality (\ref{inequa2m}),
 but the signs of the quadratic mass inequalities (\ref{Simart-r1}) and (\ref{Simart-r2}) are opposite to that of our quadratic mass inequality (\ref{inequa2m}).
   The calculations (shown in Table 3-1 and Table 3-2) manifest that
 the quadratic mass inequalities (\ref{inequa2m}) and (\ref{Simart-r0}) do satisfy the present experimental data \cite{PDG2006}
 while the quadratic mass inequalities (\ref{Simart-r1}) and (\ref{Simart-r2}) do not.
%  The inequalities (\ref{Simart-r1}) and (\ref{Simart-r2}) do not satisfy the present particle data \cite{PDG2006} (as shown in Table 2-1 and Table 2-2).

%   It should be pointed out that in the quasilinear Regge trajectory ansatz
% the additivity of intercepts and inverse slopes
% are not only suitable for the ground states
% but also for the radial and orbital excited states \cite{De-minLi2004,Li-Liu-2007}.
%    Therefore, unlike the mass inequalities extracted in
% Refs. \cite{inequalities-QCD1,inequalities-QCD2,inequalities-QCD3}
% which are only suitable for ground states,
%  the linear mass inequalities, the quadratic mass inequalities and the quadratic mass equalities
% extracted in the present work are also suitable for radial and
% orbital excited states. \\
%
%  So, unlike the linear mass inequalities extracted in Refs. \cite{inequalities-QCD1,inequalities-QCD2,inequalities-QCD3}  % which are only suitable for ground states,
% extracted in the present work are suitable for both ground and excited multiplets.

%
% We will attempt to prove them in other ways such as in QCD in the next work.
%

   We stress that quadratic baryon mass inequality (\ref{inequa2b}) has not been given before.
   From Tables 3-1, 3-2 and 4, we can see that the inequalities (\ref{inequa1m}), (\ref{inequa2m}),
(\ref{inequa1b}) and (\ref{inequa2b}) agree well with the existing experimental data \cite{PDG2006}.
   These inequalities (\ref{inequa1m}), (\ref{inequa2m}), (\ref{inequa1b}) and (\ref{inequa2b}) indicate the existence of higher-order breaking effects.

%    Eq. (\ref{high-equal-m}) is the mass relation among one $\mathcal{N} ^{2S+1}L_{J}$ multiplet. % with high accuracy.
%    Eq. (\ref{equa-double}) can be widely used to predict masses of heavy baryons when the experimental data is enough in the near future.

   For the Regge slopes of $\frac{3}{2}^+$ $SU(4)$ baryons, from Table 8, we can see that
$\alpha_{\Delta}'>\alpha_{\Sigma^*}'>\alpha_{\Xi^*}'>\alpha_{\Omega}'>  \alpha_{\Sigma_c^*}'>\alpha_{\Xi_c^*}'>\alpha_{\Omega_c^*}' >     \alpha_{\Xi_{cc}^*}'>\alpha_{\Omega_{cc}^*}'>  \alpha_{\Omega_{ccc}}'$  and
$a_{\Delta}(0)>a_{\Sigma^*}(0)>a_{\Xi^*}(0)>a_{\Omega}(0)>  a_{\Sigma_c^*}(0)>a_{\Xi_c^*}(0)>a_{\Omega_c^*}(0) >     a_{\Xi_{cc}^*}(0)>a_{\Omega_{cc}^*}(0)>  a_{\Omega_{ccc}}(0)$.
  These inequalities coincide with the expectation that the slopes of Regge trajectories decrease with quark mass increase (flavor dependent).

       From Table 2, we can see that the values of $\delta_{ij}^m$ are very sensitive to quark flavors $i$ and $j$.
  % is only a little different between different multiplets, although it is sensitive to different quark flavors $i$ or $j$.
  For the same $i$ and $j$, $ \delta_{ij}^m $ are approximately a constant (only a little different among different multiplets).
This character may be used to predict meson masses approximately in some cases.  % (flavor-dependent and multiplet-independent).
  The calculations (Table 2) show that $ \delta_{ns} < \delta_{sc} < \delta_{nc} < \delta_{cb} < \delta_{sb} < \delta_{nb} $.
  For the light mesons and baryons, $\delta_{ns}$ is close to zero.
  Letting $\delta$ $\rightarrow$ 0, one can get the usual
 Gell-Mann--Okubo quadratic relations, namely the first order of Gell-Mann--Okubo relations.
   For the heavy mesons or baryons, $\delta_{Qq}$ are large. In this case,
the quadratic mass inequalities are far from equalities.    % This also can be seen in the heavy quark effective theory (HQET).
  These features imply that the higher-order breaking effects arise with the quark mass increase.
%\vspace{1cm}
%  The linear mass inequalities and the quadratic mass inequalities are close to saturation for light mesons and charmed or bottom sextet.
% It can be believed that the mass inequalities and quadratic mass inequalities are close to saturation in $n\bar{s}$ sector.
% It can be believed that these linear mass inequalities are close to saturation in many cases.
% It can be believed that these linear mass inequalities and quadratic mass inequalities are close to saturation in many cases.

   To the second order, for baryons, as shown by Okubo long ago \cite{Okubo-2-order},
 both the well known mass relation for the baryon octet (Eq. (\ref{GMO-octet}))
 and the equal spacing rule for the baryon decuplet ($M_\Omega-M_{\Xi^*}=M_{\Xi^*}-M_{\Sigma^*}=M_{\Sigma^*}-M_{\Delta}$)
 do not hold.
   Only one relation remains,
\begin{equation}
M_\Omega-M_\Delta=3(M_{\Xi^*}-M_{\Sigma^*}) \label{2-order-Okubo-lin}.
\end{equation}
    This second-order linear mass equation was given
 by Morpurgo in the relativistic field theory \cite{decuplet-splitting}
 and by Lebed in the chiral perturbation theory \cite{decuplet-splitting2}
 and was also given in Refs. \cite{Hendry-Lichtenberg-quarkmodel,Verma-Khanna-SU4,Verma-Khanna-SU8, Singh-Verma-Khanna, Singh-Khanna, Singh} mentioned above.

 A special equation among the masses of baryons involving only two flavors % belonging to one multiplet
can be derived by taking
$\delta_{ij}^b|_{q=i}=\delta_{ij}^b|_{q=j}$ in Eq. (\ref{baryon-equal}),
\begin{equation}
\delta_{ij}^b|_{q=i}=M_{iii}^2+M_{jji}^2-2M_{iij}^2=\delta_{ij}^b|_{q=j}=M_{iij}^2+M_{jjj}^2-2M_{ijj}^2,
\end{equation}
namely,
\begin{equation}
M_{jjj}^2-M_{iii}^2=3(M_{ijj}^2-M_{iij}^2)  \label{2-order-Okubo-quadr-univ} .
\end{equation}
In the light quark sector, when $i=n$, $j=s$, for the
$\frac{3}{2}^+$ multiplet, we have
\begin{equation}
M_{\Omega}^2-M_{\Delta}^2=3(M_{\Xi^\ast}^2-M_{\Sigma^\ast}^2)
\label{2-order-Okubo-quadr}.
\end{equation}
  The quadratic equation (\ref{2-order-Okubo-quadr}) was also given by Tait
in the study of the unification $SO(6,1)$ as a spectrum generating
algebra \cite{Tait-1973}.

    In the light sector,
 both the linear mass equation, Eq. (\ref{2-order-Okubo-lin}), and the quadratic mass equation, Eq. (\ref{2-order-Okubo-quadr}),
 can be satisfied by the experimental data.
   The deviations from both of them are not more than 2$\%$.
%   Both the linear mass relation (\ref{2-order-Okubo-lin}) and the quadratic mass relation (\ref{2-order-Okubo-quadr}) satisfy HQET.

    However, generally speaking, the linear mass relation and the quadratic mass relation may not be held at the same time.
    On the other hand, the quadratic mass equation (\ref{2-order-Okubo-quadr-univ}) and the linear form of Eq. (\ref{2-order-Okubo-quadr-univ})
should give very different mass values for heavy baryons.
    The masses of the charmed and bottom particles discovered in the near future
 will numerically test which of them is realized in nature.
%  Some quadratic mass relations (\ref{baryon-equal}) such as $M_{\Omega_{ccc}}^2-M_{\Delta}^2=3(M_{\Xi_{cc}^\ast}^2-M_{\Sigma_{c}^\ast}^2)$, can be examined in the near future. % after the masses of $\Xi_{cc}^*$ and $\Omega_{ccc}$ are established in experiment.

    Theoretically, we also have some reasons besides the Regge theory to believe that
mass formulas for mesons and baryons should take the quadratic form rather than the linear form:
 1) The square of the mass operator ($M^2$) is the Casimir invariant of the Poincare group independent of any certain frame \cite{Okubo-Ryan}; % independent of any particular frame.
 2) Formulas given by asymptotic chiral symmetry are indeed in quadratic form \cite{Oneda-Matsuda-Takasugi};
 3) In the infinite-momentum frame, formulas between energy eigenvalues of hadrons spontaneously lead to quadratic mass formulas \cite{Gursey-TD Lee};
 4) Analysis on the algebraic approach indeed leads to quadratic mass formulas  \cite{Tait-1973,Oneda-Terasaki}.
     It was pointed out that the quadratic mass formula can be approximately written as the relevant linear mass formula
 when the mass splittings between the hadrons of the formula are small compared with the hadron masses \cite{Okubo-Ryan,Gursey-TD Lee}.

%$\Omega'$, $J^P=\frac{7}{2}^+ $

   %In summary
   To sum up, we conclude that quasilinear Regge trajectory and the additivity of intercepts and inverse slopes
are indeed suitable to describe meson spectra and baryon spectra at present.
 The mass relations and the predictions may
be useful for the discovery of the unobserved meson and baryon states and
the $J^P$ assignment of the meson and baryon states which will be observed in the future.

\begin{acknowledgments}
  This work was supported in part by the National Natural Science Foundation of China (Project No. 10675022),
 the Key Project of Chinese Ministry of Education (Project No. 106024)
 and the Special Grants from Beijing Normal University.
\end{acknowledgments}

% \newpage

\end{document}